\documentclass[aps,pra,twocolumn]{revtex4-2}
\usepackage{color}
\usepackage{array}
\usepackage{textcomp}
\usepackage{bm}
\usepackage{braket}
\usepackage{amsmath}
\usepackage{amssymb}
\usepackage{nccmath}
\usepackage{upgreek}
\usepackage{multirow}

\usepackage{graphicx}
\usepackage{hyperref}
\hypersetup{colorlinks=true, citecolor=blue, urlcolor=blue, linkcolor=blue}

\begin{document}

\title{Enantioselective chiral orientation induced by a combination of a long and a short laser pulse}
\author{Long Xu}
\email{long.xu@weizmann.ac.il}
\affiliation{AMOS and Department of Chemical and Biological Physics, The Weizmann Institute of Science, Rehovot 7610001, Israel}
\begin{abstract}
Enantioselective orientation of chiral molecules excited by a shaped picosecond laser pulse and a delayed femtosecond pulse is considered.
Using quantum mechanical simulations, we demonstrate a strong field-free enantioselective orientation along the laser propagation direction.
In addition, we use a classical model to reproduce the enantioselective orientation. Moreover, the analysis of the corresponding classical system allows understanding the qualitative features of the induced enantioselective orientation.
The strong enantioselective orientation may be used for the separation of chiral enantiomers using inhomogeneous electrostatic fields.
\end{abstract}
\maketitle

\section{Introduction}
Chiral molecules exist in two species --- left- and right-handed enantiomers. These enantiomers are mirror images of each other and they cannot be superimposed on each other by translations and rotations \citep{Cotton1990Chemical}.
Since its discovery by Louis Pasteur in 1848 \citep{Pasteur1848},
the phenomenon of molecular chirality has gained immense importance in biology, physics, chemistry, and the pharmaceutical industry.
Various methods for chiral discrimination have been proposed and implemented over the years. Examples involving laser fields include photoelectron circular dichroism \cite{Ritchie1976Theory,Bowering2001Asymmetry,Lux2011Circular,Beaulieu2017Attosecond,Beaulieu2018Photoexcitation}, microwave three-wave mixing spectroscopy \cite{patterson2013enantiomer,Patterson2013Sensitive,Patterson2014New,Alvin2014Enantiomer,lehmann2018theory,Ye2018Real,Ye2019Determination,leibscher2019principles}, Coulomb explosion imaging \cite{Pitzer2013Direct,Herwig2013Imaging,Fehre2019Enantioselective}, high-order harmonic generation \cite{Cireasa2015Probing}, and enantiospecific interaction with achiral magnetic substrates \cite{Banerjee-Ghosh2018}.

In the last few years, the enantioselective orientation of chiral molecules excited by laser pulses with twisted polarization has been investigated \cite{Yachmenev2016Detecting,Gershnabel2018Orienting,Tutunnikov2018Selective,Milner2019Controlled,Tutunnikov2019Laser,Tutunnikov2020Observation}.
Examples of fields with twisted polarization include the delayed cross-polarized laser pulses \cite{Fleischer2009Controlling,Kitano2009Ultrafast,Khodorkovsky2011Controlling}, the optical centrifuge for molecules \cite{Karczmarek1999Optical,Villeneuve2000Forced,Yuan2011Dynamics,Korobenko2014Direct,Korobenko2018Control}, chiral pulse trains \cite{Zhdanovich2011Control,Johannes2012Molecular}, and polarization-shaped pulses \cite{Kida2008Stimulated,Kida2009Coherent,Karras2015Polarization,Prost2017Third,Mizuse2020Direct}.
In addition, it was theoretically shown that terahertz (THz) pulses with twisted polarization \cite{Tutunnikov2021Enantioselective} and two-color laser pulses \cite{Takemoto2008Fixing, Xu2021Three} are effective for inducing enantioselective orientation.

When molecules are exposed to linearly polarized non-resonant laser fields, the most polarizable molecular axis tends to align along the polarization direction.
When molecules are excited by laser pulses with twisted polarization, the most polarizable molecular axis tends to follow the rotating polarization vector, resulting in unidirectional rotation \cite{Fleischer2009Controlling, Kitano2009Ultrafast, Khodorkovsky2011Controlling, Korech2013Observing, Mizuse2015Quantum, Lin2015Visualizing}.
Furthermore, in the case of chiral molecules, such twisted fields induce an orienting torque along the most polarizable axis \cite{Gershnabel2018Orienting,Tutunnikov2018Selective}.
This additional torque enantioselectively orients the chiral molecules perpendicular to the plane of polarization twisting. The orientation direction (along or against the laser propagation direction) depends on the sense of twisting and the handedness of the molecule.

In this paper, we study the orientation of chiral molecules excited by a linearly polarized shaped picosecond laser pulse \cite{Seideman2001On, Underwood2003Switched, Underwood2005Field, Goban2008Laser, Chatterley2019Long}
and a delayed co-propagating cross-polarized femtosecond pulse.
A slowly rising and sharply truncated picosecond pulse is used to produce a highly aligned molecular state. With long pulses, a high degree of alignment can be reached without significant molecular ionization.
Achieving the same degree of alignment with femtosecond laser pulse would require a much higher peak power, which inevitably results in sizable molecular ionization.
After the shaped picosecond pulse is suddenly switched off, a cross-polarized femtosecond laser pulse is applied, leading to enantioselective orientation along a direction perpendicular to the plane of polarization twisting, that is, along the propagation direction of the pulses.
The highly aligned state induced by the picosecond pulse enables a high degree of enantioselective orientation.
A classical model reproduces well the enantioselective orientation, and is used to analyze the various qualitative features of the effect, including the vanishing orientation along any direction in the plane of laser polarizations, the enantioselective orientation along the laser propagation direction, and the roles played by the picosecond and the femtosecond laser  pulses.

\begin{table*}[!t]\caption{Molecular properties of (\emph{R})-PPO molecule: moments of inertia (in atomic units), elements of the dipole moment (in Debye), polarizability tensors (in atomic units) expressed in the frame of principal axes of inertia.
\label{tab:Molecular-properties-PPO}}
\begin{centering}\setlength{\tabcolsep}{9mm}{
\begin{tabular}{cccc}\hline\hline
 Molecule                  & Moments of inertia            & Dipole
components           & Polarizability components                                    \\ \hline
\multirow{3}{*}{(\emph{R})-PPO} & \multirow{3}{*}{\begin{tabular}[c]{@{}c@{}}$I_{a}=180386$\\  $I_{b}=493185$\\ $I_{c}=553513$\end{tabular}} & \multirow{3}{*}{\begin{tabular}[c]{@{}c@{}}$\mu_{a}=0.965$\\$\mu_{b}=-1.733$\\$\mu_{c}= 0.489$\end{tabular}}& \multirow{3}{*}{\begin{tabular}[c]{@{}c@{}}$\alpha_{aa}=45.63$, $\alpha_{ab}=2.56$ \\ $\alpha_{bb}=37.96$, $\alpha_{ac}=0.85$\\ $\alpha_{cc}=37.87$, $\alpha_{bc}=0.65$\end{tabular} }   \\
                    &                         &                     &                   \\
                    &                         &                     &                                \\
                   \hline\hline
\end{tabular}}
\par\end{centering}
\centering{}
\end{table*}

\section{Numerical methods}

For the quantum mechanical simulations, we consider the chiral molecule as a rigid rotor.
The Hamiltonian describing the molecular rotation driven by the laser field is given by $H(t)=H_{r}+H_{\mathrm{int}}(t)$ \citep{Krems2018Molecules,Koch2019Quantum}, where $H_{r}$ is the field-free Hamiltonian and $H_{\mathrm{int}}(t)=-\mathbf{E}\cdot(\bm{\alpha}\mathbf{E})/2$ is the interaction of molecule with the laser field.
Here $\bm{\alpha}$ is the molecular polarizability tensor and $\mathbf{E}$ is the external electric field.
The wave function is expressed in the basis of field-free symmetric-top eigenfunctions $|JKM\rangle$ \citep{zare1988Angular}.
Here $J$ is the total angular momentum, while $K$ and $M$ are its projections on the molecule-fixed $z$ and the laboratory-fixed $Z$ axes respectively.
The nonzero matrix elements of the field-free Hamiltonian are \citep{zare1988Angular}
\begin{flalign}
&\langle JKM|H_{r}|JKM\rangle=\frac{B+C}{2} \left[J(J+1)-K^{2}\right]+AK^{2},\nonumber\\
&\langle JKM|H_{r}|J,K\pm2,M\rangle  =\frac{B-C}{4}f(J,K\pm1),\label{eq:HR2}
\end{flalign}
where $f(J,K)=\sqrt{(J^{2}-K^{2})[(J+1)^{2}-K^{2}]}$, $A=\hbar^{2}/2I_{a}$, $B=\hbar^{2}/2I_{b}$, $C=\hbar^{2}/2I_{c}$ with the moments of inertia $I_{a}<I_{b}<I_{c}$.
The initial state is the ground state, corresponding to a rotational temperature of $0\,\mathrm{K}$.
The time-dependent Schr\"{o}dinger equation $i\hbar\partial_{t}|\Psi(t)\rangle=H(t)|\Psi(t)\rangle$ is solved by numerical exponentiation of the Hamiltonian matrix (see Expokit \citep{sidje1998Expokit}).
A detailed description of our numerical scheme can be found in \citep{Tutunnikov2019Laser}.

\section{strong enantioselective orientation}

\begin{figure}[!b]
\centering{}
\includegraphics[width=\linewidth]{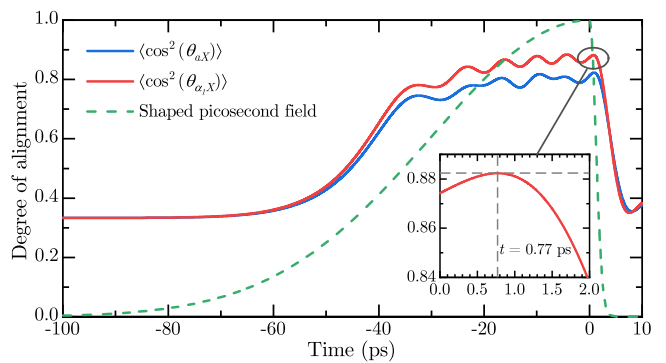}
\caption{Quantum mechanically calculated time-dependent degree of molecular alignment along the polarization direction of the picosecond pulse.
The initial temperature is $0\,\mathrm{K}$. Note that the degrees of alignment for the two enantiomers, (\emph{R})- and (\emph{S})-PPO are the same.
The green dashed line represents the envelope of the shaped picosecond field [see Eq. \eqref{eq:adiabatic_laser}]. The inset shows the magnified portion of $\braket{\cos^2\,(\theta_{\alpha_1 X})}$.
 \label{fig:alignment}}
\end{figure}

Here we consider propylene oxide molecule (PPO, $\mathrm{CH_{3}CHCH_{2}O}$) as a typical example of a chiral molecule.
Table \ref{tab:Molecular-properties-PPO} summarizes the molecular properties of (\emph{R})-PPO (right-handed enantiomer), which were computed using GAUSSIAN software package (method: CAM-B3LYP/aug-cc-pVTZ) \cite{Frisch2016Gaussian}.
The molecules are excited by a shaped picosecond laser pulse followed by a femtosecond laser pulse. Both pulses propagate in the same direction, the $Z$ direction.
The picosecond laser pulse is polarized along the laboratory-fixed $X$ axis and it is used to produce a highly aligned molecular state, while the femtosecond laser pulse, polarized at angle $\pi/4$ to the polarization of the picosecond pulse, is used to induce the enantioselective orientation.
The electric field of the shaped picosecond laser pulse (see Fig. \ref{fig:alignment}) is given by
\begin{equation}
    \mathbf{E}^p(t)=E^p_0\exp\left(-2\ln 2 \frac{t^2}{\sigma^2}\right)\cos(\omega t)\mathbf{e}_X, \label{eq:adiabatic_laser}
\end{equation}
where
\begin{equation*}
    \begin{cases}
\sigma=\sigma_\mathrm{on},& t<0,\\
\sigma=\sigma_\mathrm{off},& t\geq 0.
\end{cases}
\end{equation*}
The electric field of the femtosecond laser pulse is
\begin{fleqn}
\begin{eqnarray}
    \mathbf{E}^f(t)=
\frac{E^f_0}{\sqrt{2}}\exp\left[-2\ln 2 \frac{(t-\tau)^2}{\sigma_f^2}\right]\cos(\omega t)(\mathbf{e}_X+\mathbf{e}_Y). \label{eq:femtosecond_laser}
\end{eqnarray}
\end{fleqn}
Here $E^p_0 \,(E^f_0)$ is the peak amplitude of the picosecond (femtosecond) field, $\omega$ is the carrier frequency, and $\mathbf{e}_X \,(\mathbf{e}_Y)$ is a unit vector along the laboratory $X \,(Y)$ axis.
$\sigma_\mathrm{on}$ and $\sigma_\mathrm{off}$ represent the full width at half maximum (FWHM) of the slow switch-on and the rapid switch-off, respectively.
$\sigma_f$ is the FWHM of the femtosecond laser pulse and $\tau$ is defined as the time at which the femtosecond laser pulse is applied.

Figure \ref{fig:alignment} shows the time-dependent degree of alignment, quantified by the average of the squares of the directional cosines, $\braket{\cos^2\,(\theta_{aX})}(t)=\langle \Psi(t)|(\bm{a}\cdot \mathbf{e}_X)^2|\Psi(t)\rangle$ and $\braket{\cos^2\,(\theta_{\alpha_1 X})}(t)=\langle \Psi(t)|(\bm{\alpha}_1\cdot \mathbf{e}_X)^2|\Psi(t)\rangle$.
$\bm{a}=(0,0,1)$ is the molecule-fixed $a$ axis, while $\bm{\alpha}_1$ is the most polarizable molecular axis and for (\emph{R})-PPO molecule, $\bm{\alpha}_1 = (0.293, 0.115, 0.949)$.
The vectors are expressed in the molecule-fixed frame using the basis consisting of the three inertia principal axes $b,\,c,\,a$, namely $x\rightarrow b$, $y\rightarrow c$, $z\rightarrow a$, such that an arbitrary vector $\bm{r}$ has the Cartesian components $\bm{r}=(r_b,r_c,r_a)$ in the molecule-fixed frame.
$\theta_{aX}$ ($\theta_{\alpha_1 X}$) denotes the angle between the molecule-fixed $a$ ($\alpha_1$) axis and the laboratory-fixed
$X$ axis (polarization direction of the picosecond pulse).
Here the initial state is the ground rotational state (temperature is $0\,\mathrm{K}$).
The peak intensity of the picosecond laser pulse is $I^p=10^{12} \,\mathrm{W/cm^2}$, the FWHMs are $\sigma_\mathrm{on}=50\,\mathrm{ps}$ and $\sigma_\mathrm{off}=2\,\mathrm{ps}$.
In the presence of the picosecond pulse, the most polarizable molecular $\alpha_1$ axis is aligned along the polarization direction.
Figure \ref{fig:alignment} shows high degrees of alignment of the molecular $a$ and $\alpha_1$ axes building up during the picosecond laser pulse.
The maximal degrees of alignment are $\braket{\cos^2\,(\theta_{a X})}\simeq 0.82$ and $\braket{\cos^2\,(\theta_{\alpha_1 X})}\simeq 0.88$.

\begin{figure}[!t]
\centering{}
\includegraphics[width=\linewidth]{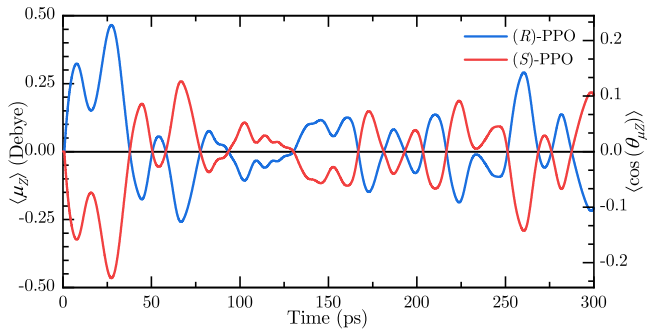}
\caption{Quantum expectation value of $Z$ projection of the dipole signal, $\braket{\mu_Z}(t)$ or the degree of orientation, $\braket{\cos\,(\theta_{\mu Z})}\equiv\braket{\mu_Z}(t)/\mu$.
Here $\theta_{\mu Z}$ is the angle between the molecular dipole moment and the laboratory $Z$ axis, $\mu$ is the magnitude of the dipole moment.
(\emph{R})- and (\emph{S})-PPO denote the right- and left-handed enantiomers of the PPO molecule, respectively.
Note that the signals are $\pi$ out of phase for the two enantiomers.
 \label{fig:Quantum}}
\end{figure}

During the turn-off stage of the shaped picosecond pulse, around the time when  the degree of alignment at maximum (see the inset in Fig. \ref{fig:alignment}), a cross-polarized femtosecond laser pulse is applied. The femtosecond laser pulse is polarized at angle $\pi/4$ to the $X$ axis.
The peak intensity of the femtosecond laser pulse is $I^f=10^{13} \,\mathrm{W/cm^2}$, its FWHM is $\sigma_f=0.1\,\mathrm{ps}$, and $\tau=0.77\,\mathrm{ps}$, see Eq. \eqref{eq:femtosecond_laser}.
Figure \ref{fig:Quantum} shows the time-dependent dipole signal along the laboratory $Z$ axis, $\braket{\mu_Z}(t)=\langle \Psi(t)|\bm{\mu}\cdot \mathbf{e}_Z|\Psi(t)\rangle$ ($\bm{\mu}$ is the permanent dipole moment).
As expected, shortly after the pulse, enantioselective orientation along the laser propagation direction appears, namely the sign of $\braket{\mu_Z}(t)$ is positive (negative) for (\emph{R})-PPO and negative (positive) for (\emph{S})-PPO.
The degree of orientation along any axis in the $XY$ plane is identically zero.
The general behavior is consistent with previous studies on the enantioselective orientation induced by laser pulses with twisted polarization \cite{Yachmenev2016Detecting,Gershnabel2018Orienting,Tutunnikov2018Selective,Milner2019Controlled,Tutunnikov2019Laser,Tutunnikov2020Observation}.
The scheme presented here, which is based on the shaped picosecond and femtosecond laser pulses, achieves comparable degree of orientation, resulting in a field-free transient dipole signal of about 0.47 Debye corresponding to a degree of orientation $\braket{\cos\,(\theta_{\mu Z})}\simeq 0.23$.
The high degree of alignment generated by the picosecond field (see Fig. \ref{fig:alignment}) serves as a basis for inducing the high degree of enantioselective orientation using the femtosecond laser pulse  polarized at angle $\pi/4$ to the polarization direction of the picosecond pulse.\\

\section{Qualitative description --- Classical model}

To understand the origin of the enantioselective orientation shown in Fig. \ref{fig:Quantum} and the role of each of the laser pulses, we resort to a classical model. We consider a collection of classical non-interacting chiral molecules, where each molecule is modeled by a rigid polarizable asymmetric-top.
The picosecond field prepares a highly aligned molecular ensemble, where the most polarizable molecular $\alpha_1$ axis points, on average, along the field polarization direction.
To simplify the analysis, we assume that initially (immediately after the end of the picosecond pulse)  all the molecular $a$ axes (which are close to the $\alpha_1$ axis) are perfectly aligned along the laboratory $X$ axis, and we take the initial temperature to be $0\,\mathrm{K}$.
The aligned molecular ensemble is then excited, at $t=\tau=0.77\,\mathrm{ps}$, by a femtosecond laser pulse polarized at angle $\pi/4$ to the $X$ axis (in the $XY$ plane), see Eq. \eqref{eq:femtosecond_laser}.

\begin{figure}[!t]
\centering{}
\includegraphics[width=\linewidth]{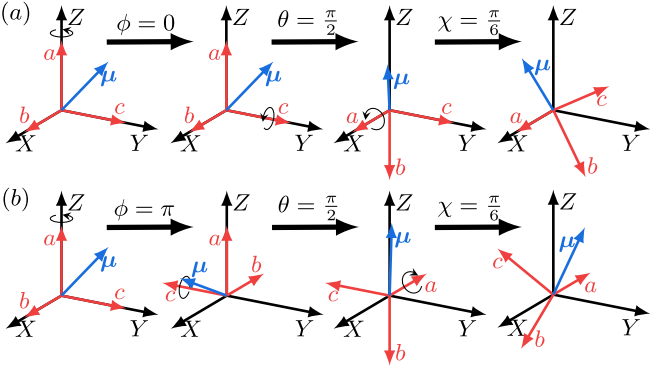}
\caption{A set of three Euler rotations (in accordance with the convention of \cite{zare1988Angular}) describing the orientation of the molecule-fixed frame ($bca$) relative to the laboratory-fixed frame ($XYZ$) for (a) $A=(0,\pi/2,\pi/6)$ and (b) $B=(\pi,\pi/2,\pi/6)$.
The blue arrows represent the dipole moment ($\bm{\mu}$) of a (\emph{R})-PPO molecule.
 \label{fig:Euler_angles}}
 \end{figure}

Classically, the rotation of a rigid body is described by Euler's equations \citep{Goldstein2002Classical}
\begin{equation}
\mathbf{I}\bm{\dot{\Omega}}=(\mathbf{I}\bm{\Omega})\times\bm{\Omega}+\mathbf{T},\label{eq:Eulers-equations}
\end{equation}
where $\bm{\Omega}=(\Omega_{b},\Omega_{c},\Omega_{a})$ is the angular velocity vector, $\mathbf{I}=\mathrm{diag}(I_{b},I_{c},I_{a})$ is the moment of inertia tensor, and $\mathbf{T}=(T_{b},T_{c},T_{a})$ is the external torque vector.
All the quantities in Eq. \eqref{eq:Eulers-equations} are expressed in the molecule-fixed frame of reference using the basis consisting of the three principal axes of inertia, $b,\,c,\,a$.
The torque is given by $\mathbf{T} = \boldsymbol{\alpha}\mathbf{E}\times\mathbf{E}$ and it can be obtained by expressing the electric field vector in the molecule-fixed frame.
The relation between the laboratory-fixed and the molecule-fixed frames is described by $(X, Y, Z)^T=U(\phi,\theta,\chi)(b, c, a)^T$, where the transformation matrix is given by \cite{zare1988Angular}
\begin{widetext}
\begin{align}\label{eq:transformation}
    U(A)
    =\begin{pmatrix}
    \cos(\phi) \cos(\theta) \cos(\chi) - \sin(\phi) \sin(\chi) & -\cos(\phi) \cos(\theta) \sin(\chi) - \sin(\phi) \cos(\chi) & \cos(\phi) \sin(\theta)\\
    \sin(\phi) \cos(\theta) \cos(\chi) + \cos(\phi) \sin(\chi) & -\sin(\phi) \cos(\theta) \sin(\chi) + \cos(\phi) \cos(\chi) & \sin(\phi) \sin(\theta)\\
    -\sin(\theta) \cos(\chi) & \sin(\theta) \sin(\chi) & \cos(\theta)
  \end{pmatrix}.
\end{align}
\end{widetext}
Here $A=(\phi,\theta,\chi)$ is a shorthand notation for the three Euler angles: $\phi$ and $\theta$ are the azimuthal and polar angles defining the orientation of the molecular $z$ axis (associated with the molecular $a$ axis here) in the laboratory-fixed frame, and $\chi$ is the additional rotation angle about the $z$ axis, see Fig. \ref{fig:Euler_angles}.
The rates of change of the Euler angles are given by \cite{zare1988Angular}
\begin{subequations}
\begin{align}
\dot{\phi}&=\frac{- \Omega_{b}\cos (\chi)+\Omega_{c}\sin (\chi) }{\sin (\theta)}, \label{eq:angle_a}\\
\dot{\theta}&=\Omega_{b}\sin (\chi) +\Omega_{c}\cos (\chi), \label{eq:angle_b}\\
\dot{\chi}&=\Omega_{a}-\cos (\theta) \frac{-\Omega_{b}\cos (\chi) + \Omega_{c}\sin (\chi) }{\sin (\theta)}. \label{eq:angle_c}
\end{align}\end{subequations}

To describe the ensemble behavior, we use the Monte Carlo approach and consider the dynamics of $N\gg 1$ chiral molecules.
For each molecule, we numerically solve a system of differential equations consisting of Eqs. \eqref{eq:Eulers-equations} and (\ref{eq:angle_a}-\ref{eq:angle_c}).
The $N$ molecules are divided into two groups, I and II. The initial state of the molecules in group I is given by $A(t_0)=(0,\pi/2,\chi_0)$, where $\chi_0$ is uniformly distributed between 0 and $2\pi$, corresponding to the initial state in which the molecular $a$ axes point \emph{along} the laboratory $X$ axis [see Fig. \ref{fig:Euler_angles}(a)]. The initial state of the molecules in group II is defined by $
B(t_0)=(\pi,\pi/2,\chi_0)$, such that the molecular $a$ axes point \emph{against} the laboratory $X$ axis [see Fig. \ref{fig:Euler_angles}(b)].
The initial angular velocities are set to $\Omega_i=0$, $i=b,c,a$, corresponding to a temperature of $0\,\mathrm{K}$.

\begin{figure}[!t]
\centering{}
\includegraphics[width=\linewidth]{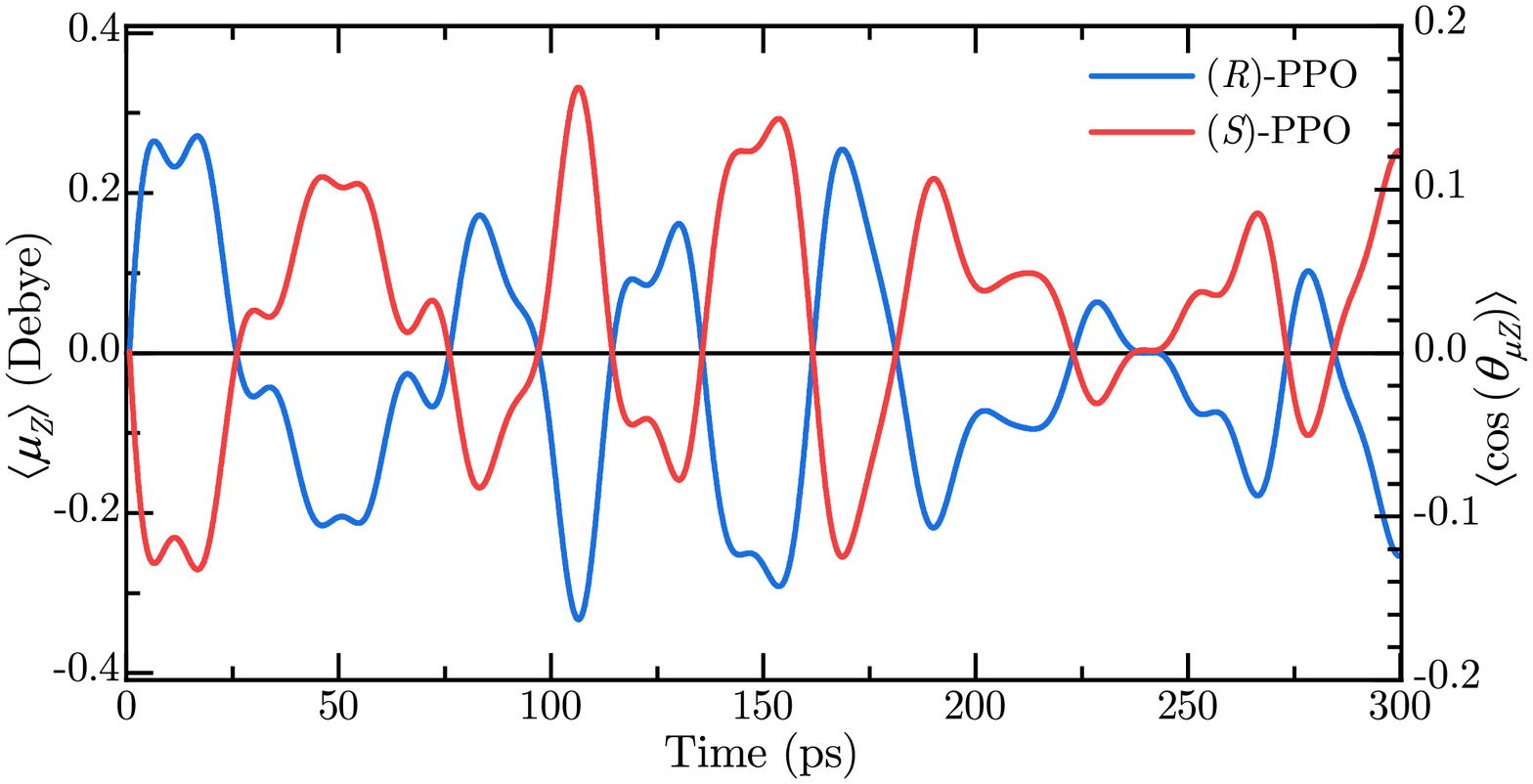}
\caption{Classically calculated $Z$ projection of the dipole signal or the degree of orientation. Initially, the molecular $a$ axes are aligned along the $X$ axis, and then the aligned molecules are excited by a femtosecond laser pulse polarized at an angle $\pi/4$ to the $X$ axis.
The initial temperature is $0\,\mathrm{K}$. The field parameters of the femtosecond laser pulse are the same as the case shown in Fig. \ref{fig:Quantum}.
$N = 10^7$ molecules are used in the classical simulation.
\label{fig:Classical_a}}
\end{figure}

For convenience, the femtosecond laser pulse is treated as an impulsive excitation, such that the angular velocity just after the pulse is directly proportional to the applied torque, $\Omega_i \propto T_i/I_i$.
Considering two molecules, one from each group, in initial states $A(t_0)$ and $B(t_0)$, the torque components in the rotating molecular frame are the same for the two molecules. After averaging over the rapid oscillations of the laser field (the carrier frequency, $\omega$ is larger than the typical rotation frequency of the molecule by several orders of magnitude), the torque components are given by
\begin{fleqn}
\begin{subequations}
\begin{align}
    T_b (\chi_0)&= \frac{(E^f)^2}{4}
    \Big[(\alpha_{cc}-\alpha_{aa})\cos(\chi_0)+\alpha_{bc}\sin(\chi_0)\nonumber\\
    &+\alpha_{ac}\sin^2(\chi_0)-\alpha_{ab}\sin(\chi_0)\cos(\chi_0)\Big], \label{eq:torques_a}\\
    T_c (\chi_0)&= \frac{(E^f)^2}{4}
    \Big[(\alpha_{aa}-\alpha_{bb})\sin(\chi_0)
   -\alpha_{bc}\cos(\chi_0)   \nonumber\\
    & -\alpha_{ab}\cos^2(\chi_0) +\alpha_{ac}\sin(\chi_0)\cos(\chi_0)\Big], \label{eq:torques_b}\\
    T_a (\chi_0)&= \frac{(E^f)^2}{4}
    \Big[  (\alpha_{bb}-\alpha_{cc})\sin(\chi_0)\cos(\chi_0)
    \nonumber\\
    &+\alpha_{bc}\cos(2\chi_0)
    +\alpha_{ab}\cos(\chi_0)-\alpha_{ac}\sin(\chi_0)\Big]. \! \! \label{eq:torques_c}
\end{align}
\end{subequations}
\end{fleqn}
Figure \ref{fig:Classical_a} shows the classically calculated ensemble average, $\braket{\mu_Z}(t)$.
Here, the initial conditions and the field parameters of the femtosecond pulse are the same as in Fig. \ref{fig:Quantum}. Comparing Figs. \ref{fig:Quantum} and \ref{fig:Classical_a}, it is evident that the simplified classical model qualitatively reproduces the enantioselective orientation. Therefore, in what follows, we analyze the system using the classical model in order to gain physical insight to the qualitative features of the enantioselective orientation.

\subsection{Absence of orientation in the plane of the pulses’ polarizations --- \texorpdfstring{$XY$}{} plane}\label{sec:XYdirections}

In this subsection, we consider the orientation along the laser propagation direction and the absence thereof of orientation in any perpendicular direction. Here, we focus on a single enantiomer. The next subsection addresses the enantioselectivity. As can be seen from Eqs. (\ref{eq:angle_a}-\ref{eq:angle_c}) and Eqs. (\ref{eq:torques_a}-\ref{eq:torques_c}), the torques (the induced angular velocities) and the rate of change of the Euler angles are independent of $\phi$.
Accordingly, the initial states $A(t_0)$ and $B(t_0)$ evolve into $A(t)=(\delta \phi, \delta \theta +\pi/2, \delta\chi+\chi_0)$ and $B(t)=(\delta \phi+\pi, \delta \theta +\pi/2, \delta\chi+\chi_0)$, respectively. Here $\delta \phi$, $\delta \theta$, and $\delta \chi$ are the changes in Euler angles resulting from the rotational dynamics induced by the femtosecond laser pulse kick. Notice the $\pi$ difference between $\phi$ angles in $A(t)$ and $B(t)$.

The time-dependent projections of the dipole moment on the laboratory axes can be obtained using the transformation matrix $U$ [see Eq. \eqref{eq:transformation}]
\begin{subequations}
\begin{align}
U\left[A(t)\right](\mu_b,\mu_c,\mu_a)^T &=(\mu_X,\mu_Y,\mu_Z)^T,  \label{eq:A(t)}
\\ U\left[B(t)\right](\mu_b,\mu_c,\mu_a)^T &=(-\mu_X,-\mu_Y,\mu_Z)^T. \label{eq:B(t)}
\end{align}
\end{subequations}
Equations \eqref{eq:A(t)} and \eqref{eq:B(t)} follow from the fact that the initial $\pi$ difference in the $\phi$ angles of $A$ and $B$ is preserved throughout the motion.
The same relation can also be found in Fig. \ref{fig:distribution-RPPO}, where we plot the dipole signals as functions of $\chi_0$ for the two molecular groups ($\phi_0=0$ and $\phi_0=\pi$). The dipole moment projections on the $X$ and $Y$ axes are of opposite sign, and when averaged over the entire molecular ensemble (groups I and II), resulting in zero averaged dipole signal along any direction in the $XY$ plane, including $\braket{\mu_X}=\braket{\mu_Y}=0$.
In contrast, the dipole moment projections on the $Z$ axis are of the same sign, giving rise to the nonzero ensemble-averaged orientation for chiral molecules, while for non-chiral molecules, $\braket{\mu_Z}=0$ (see below for further details).

\subsection{Enantioselective orientation along the laser propagation direction --- \texorpdfstring{$Z$}{} direction} \label{sec:enantioselective}

\begin{figure}[!h]
\centering{}
\includegraphics[width=\linewidth]{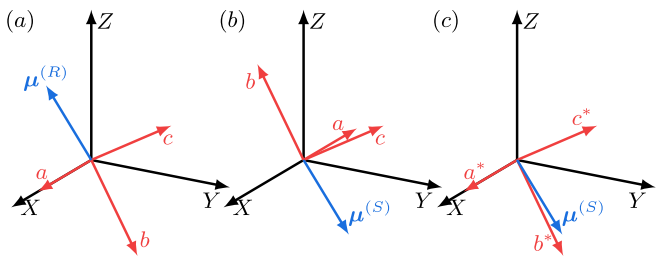}
\caption{Two initial orientations of the molecular
frame for (a) $A=(0,\pi/2,\chi_0)$ and (b) $B_1=(\pi,\pi/2,\pi-\chi_0)$. (c) Same as panel (b) for $B_1^*=(0,\pi/2,\chi_0)$, but in terms of the basis $(b^*,c^*,a^*)=(-b,c,-a)$. Here, $\chi_0=\pi/6$. $\bm{\mu}^{(R)}$ and $\bm{\mu}^{(S)}$ denote dipole moments of (\emph{R})- and (\emph{S})-PPO molecules, respectively.
Note that $\bm{\mu}^{(S)}$ in (b) and (c) point in the same directions, while $\bm{\mu}^{(R)}$ in (a) points in the opposite direction.
} \label{fig:orientation_euler_angles}
\end{figure}

In this subsection, we consider the implications of chirality on molecular orientation. The two enantiomers are related by a reflection transformation  \cite{Cotton1990Chemical}, here we set the $ab$-plane as the reflection plane. Accordingly, the dipole moment components and the polarizability components of the two enantiomers [(\emph{S}) and (\emph{R})] are related as $\mu_c^{(S)}=-\mu_c^{(R)}$,  $\alpha_{ac}^{(S)}=-\alpha_{ac}^{(R)}$, and $\alpha_{bc}^{(S)}=-\alpha_{bc}^{(R)}$. The other components are the same for both enantiomers.

\begin{figure}[!h]
\centering{}
\includegraphics[width=\linewidth]{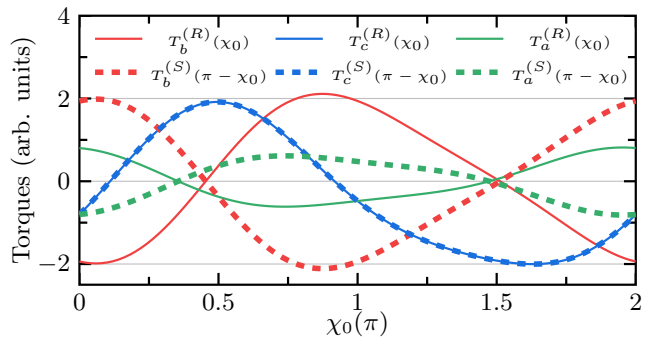}
\caption{Torque components [Eqs. (\ref{eq:torques_a}-\ref{eq:torques_c})] as functions of $\chi_0$ for (\emph{R})-PPO (solid) and (\emph{S})-PPO (dashed) molecules with $E^f = 1$.} \label{fig:torques_PPO}
\end{figure}

To analyze the enantioselective orientation, we begin with two enantiomers of the propylene oxide molecule, (\emph{R})- and (\emph{S})-PPO. Initially, the molecular $a$ axes are aligned along the $X$ axis, while the dipole moments of the two enantiomers point in the opposite directions (see Fig. \ref{fig:orientation_euler_angles}).
Without loss of generality, we consider the initial states $A(t_0)=(0,\pi/2,\chi_0)$ [panel (a)] and $B_1(t_0)=(\pi,\pi/2,\pi-\chi_0)$ [panel (b)] for (\emph{R})- and (\emph{S})-PPO, respectively. For these two states, $\boldsymbol{\mu}^{(R)}=-\boldsymbol{\mu}^{(S)}$.
Figure \ref{fig:torques_PPO} shows the torque components [see Eqs. (\ref{eq:torques_a}-\ref{eq:torques_c})] as functions of $\chi_0$ for the two enantiomers.
The relation between the torque components is given by
\begin{equation}\label{eq:chiral_RS}
T_i^{(S)}(\pi-\chi_0)=(-1)^{\delta_{ci}+1}T_i^{(R)}(\chi_0),\;i= b,\, c,\, a.
\end{equation}

To simplify the comparison of the two enantiomers, we introduce a new basis in the molecular frame of $(S)$-PPO, consisting of the $b^*,c^*,a^*$ axes, where $(b^*,c^*,a^*)=C_2(c)(b,c,a)=(-b,c,-a)$ and $C_2(c)$ is the two-fold rotation about the $c$ axis \cite{zare1988Angular}.
In terms of the new basis, the molecular orientation in Fig. \ref{fig:orientation_euler_angles}(b) is given by $B_1^*(t_0)=(0,\pi/2,\chi_0)$ [see Fig. \ref{fig:orientation_euler_angles}(c)]. Notice, $B_1^*(t_0)=A(t_0)$.
The relation between the torque components of $(R)$-PPO and $(S)$-PPO (in terms of the new basis) is
\begin{equation}
T_{i^*}^{(S)}(\chi_0)=T_i^{(R)}(\chi_0),
\end{equation}
because $T_{i^*}^{(S)}(\chi_0)=(-1)^{\delta_{ci}+1}T_i^{(S)}(\pi-\chi_0)$, $i^*=b^*,\, c^*, \,a^* $.

Thus, in this new basis, $(b^*,c^*,a^*)$, the torque components \emph{and} the initial states for both enantiomers in Fig. \ref{fig:orientation_euler_angles} are the \emph{same}. As a consequence, $B_1^{*(S)}(t)=A^{(R)}(t)$ [see Eqs. (\ref{eq:angle_a}-\ref{eq:angle_c})].
Moreover, since
$\left(\mu_{b^*}^{(S)},\mu_{c^*}^{(S)},\mu_{a^*}^{(S)}\right)=\left(-\mu_{b}^{(S)},\mu_{c}^{(S)},-\mu_{a}^{(S)}\right)=-\left(\mu_b^{(R)},\mu_c^{(R)},\mu_a^{(R)}\right)$, we obtain the relations
\begin{widetext}\begin{flalign}\label{eq:chiral}
\begin{pmatrix}
    \begin{matrix}
    \mu_X^{(S)}[B_1(t)] \\
    \mu_Y^{(S)}[B_1(t)]\\
    \mu_Z^{(S)}[B_1(t)]
    \end{matrix}
  \end{pmatrix}
 = U \left[B_1^{(S)}(t)\right]\begin{pmatrix}
    \begin{matrix}
    \mu_{b}^{(S)} \\
    \mu_{c}^{(S)}\\
    \mu_{a}^{(S)}
    \end{matrix}
  \end{pmatrix}
  = U \left[B_1^{*(S)}(t)\right]\begin{pmatrix}
    \begin{matrix}
    \mu_{b^*}^{(S)} \\
    \mu_{c^*}^{(S)}\\
    \mu_{a^*}^{(S)}
    \end{matrix}
  \end{pmatrix}= -U \left[A^{(R)}(t)\right]\begin{pmatrix}
    \begin{matrix}
    \mu_{b}^{(R)} \\
    \mu_{c}^{(R)}\\
    \mu_{a}^{(R)}
    \end{matrix}
  \end{pmatrix}
  =-\begin{pmatrix}
  \begin{matrix}
    \mu_X^{(R)}[A(t)] \\
    \mu_Y^{(R)}[A(t)]\\
    \mu_Z^{(R)}[A(t)]
    \end{matrix}
  \end{pmatrix},\!\!\!\!\!
   \end{flalign}
\end{widetext}
which indicates that the opposite sign between $\bm{\mu}^{(S)}$ and $\bm{\mu}^{(R)}$ is preserved throughout the motion (which can also be seen from Figs. \ref{fig:distribution-RPPO} and \ref{fig:distribution-SPPO}).
As discussed above in Sec. \ref{sec:XYdirections}, the ensemble-averaged dipole signal of each enantiomer along any direction in the $XY$ plane is zero.
Consequently, the orientation appears in the $Z$ direction and it is \emph{enantioselective}. Note, however, that in the case of a racemic mixture (which has equal amounts of left- and right-handed enantiomers), the orientation in the $Z$ direction vanishes.
Similar and simplified arguments were previously used in \cite{Gershnabel2018Orienting,Tutunnikov2018Selective} to explain the enantioselective orientation.

\subsection{Chirality dependence of the orientation along the laser propagation direction -- \texorpdfstring{$Z$}{} direction}

In the special case of non-chiral molecules, having diagonal polarizability tensor, the torque components in Eqs. (\ref{eq:torques_a}-\ref{eq:torques_c}) reduce to
\begin{subequations}
\begin{align}
    T_b (\chi_0)&= \frac{(E^f)^2}{4}
    (\alpha_{cc}-\alpha_{aa})\cos(\chi_0), \label{eq:torques_nonchiral_a}\\
    T_c (\chi_0)&= \frac{(E^f)^2}{4}
    (\alpha_{aa}-\alpha_{bb})\sin(\chi_0), \label{eq:torques_nonchiral_b}\\
    T_a (\chi_0)&= \frac{(E^f)^2}{4}
    (\alpha_{bb}-\alpha_{cc})\sin(\chi_0)\cos(\chi_0). \label{eq:torques_nonchiral_c}
\end{align}
\end{subequations}
In contrast to the case of chiral molecules (see Fig. \ref{fig:torques_PPO}), here the torque components have a well defined symmetry (symmetric/antisymmetric) about $\chi_0=n\pi/2$, $n=0,1,2,3$. For instance,
\begin{subequations}
\begin{align}
    T_{i}(-\chi_0)=(-1)^{\delta_{bi}+1}T_i(\chi_0), \label{eq:nonchiral_b}\\
    T_{i}(\pi-\chi_0)=(-1)^{\delta_{ci}+1}T_i(\chi_0), \label{eq:nonchiral_c}\\
    T_{i}(\pi+\chi_0)=(-1)^{\delta_{ai}+1}T_i(\chi_0). \label{eq:nonchiral_a}
\end{align}
\end{subequations}

Note that the relation in Eq. \eqref{eq:nonchiral_c} is the same as in Eq. \eqref{eq:chiral_RS}. When the dipole moment projection on the molecular $c$ axis is zero ($\mu_c=0$), all the expressions from  Sec. \ref{sec:enantioselective} still apply.
Similar to Eq. \eqref{eq:chiral}, the relation between the dipole moments corresponding to the initial states of $A(t_0)$ and $B_1(t_0)$ is given by
\begin{align}\label{eq:non_chiral}
\begin{pmatrix}
    \begin{matrix}
    \mu_X[B_1(t)] \\
    \mu_Y[B_1(t)]\\
    \mu_Z[B_1(t)]
    \end{matrix}
  \end{pmatrix}
  =-\begin{pmatrix}
  \begin{matrix}
    \mu_X[A(t)] \\
    \mu_Y[A(t)]\\
    \mu_Z[A(t)]
    \end{matrix}
  \end{pmatrix}.
   \end{align}
Consequently, the ensemble-averaged dipole signal along any laboratory axis is zero, and this includes $\braket{\mu_Z}$ the projection along the laser propagation direction.
Compared to the two enantiomers in Sec. \ref{sec:enantioselective}, the two non-chiral molecules in initial states $A(t_0)$ and $B(t_0)$ are indistinguishable and can be viewed as a mixture containing equal amounts of left- and right-handed enantiomers, leading to the vanishing of the ensemble-averaged orientation.
Moreover, notice that $ \mu_Z[B_1(t)]=-\mu_Z[A(t)]=-\mu_Z[B(t)]$ [see Eqs. \eqref{eq:A(t)} and \eqref{eq:B(t)}], such that even within each group, $A(t_0)$ or $B(t_0)$, the dipole signal $\braket{\mu_Z}$ vanishes.
The symmetries of the torque components [see Eqs. \eqref{eq:nonchiral_b} and \eqref{eq:nonchiral_a}] allow repeating the same arguments for molecules in which $\mu_b=0$ or $\mu_a=0$.

In contrast, in the case of a single enantiomer of chiral molecules, the torque components do not have symmetry similar to those in Eqs. (\ref{eq:nonchiral_b}-\ref{eq:nonchiral_a}),
resulting in the nonzero orientation along the $Z$ axis. This orientation depends on the polarizability tensor being non-diagonal, which is the case for chiral molecules.

\subsection{Roles of the laser pulses}\label{sec:roles}

For a collection of isotropically distributed chiral molecules, collinearly polarized one-color laser pulses can only induce molecular alignment, but no orientation. Furthermore, orthogonally polarized laser pulses which do not overlap cannot induce orientation either.
The reason is that the field-polarizability interaction and the spatial distribution of the molecular ensemble are always symmetric about the axis parallel/perpendicular to the polarization direction.

For the sake of simplicity, we consider an ensemble with the molecular $a$ axes initially aligned along the $X$ axis ($\phi_0=0,\pi$, and $\theta_0=\pi/2$) and being excited by a $Y$-polarized laser pulse.
In this case, the torque components denoted as $T_i'$ can be written as
\begin{align*}
    T_b' (\chi_0)&= -\frac{(E^f)^2}{2}
    \big[\alpha_{ab}\sin(\chi_0)\cos(\chi_0)+\alpha_{ac}\cos^2(\chi_0)\big], \\
    T_c' (\chi_0)&= \frac{(E^f)^2}{2}
    \big[\alpha_{ab}\sin^2(\chi_0)+\alpha_{ac}\sin(\chi_0)\cos(\chi_0)\big],  \\
    T_a' (\chi_0)&= \frac{(E^f)^2}{4}
    \big[2\alpha_{bc}\cos(2\chi_0)+(\alpha_{bb}-\alpha_{cc})\sin(2\chi_0)\big].
\end{align*}
Accordingly, $T_i'(\pi+\chi_0)=T_i'(\chi_0)$ and shortly after the pulse, $\Omega_i'(\pi+\chi_0)=\Omega_i'(\chi_0)$.
From Eqs. (\ref{eq:angle_a}-\ref{eq:angle_c}), it follows that for the same $\theta_0$,
\begin{align*}
\delta\theta'(\pi+\chi_0)=-\delta\theta'(\chi_0), \quad \delta\chi'(\pi+\chi_0)&=\delta\chi'(\chi_0),
\end{align*}
such that $\mu_{Z}'(\pi+\chi_0)=-\mu_{Z}'(\chi_0)$ [see Eq. \eqref{eq:transformation}].
As a result, because the contributions from the molecules at $\pi+\chi_0$ and $\chi_0$ cancel each other (can also be seen from Fig. \ref{fig:distribution-RPPO-Ypolarized}), the ensemble-averaged dipole signal along the $Z$ axis is zero. Since the torque components depend on $\chi_0$ only, the same argument as in Sec. \ref{sec:XYdirections} can also be applied, meaning that the dipole signal along any direction in the $XY$ plane vanishes. Thus, there is no orientation in this case.

In contrast, in the case of a pair of delayed crossed-polarized pulses, when the second laser pulse is polarized at an angle ($\neq n\pi/2, \, n=0,1,2,3$) with respect to the first one, the first pulse effectively breaks the symmetry of spatial distribution about the polarization direction of the second one.
Consequently, the induced asymmetric torques [see Eqs. (\ref{eq:torques_a}-\ref{eq:torques_c})] induce molecular orientation along the direction perpendicular to the polarization of the two exciting pulses.
It is required that the time scales of polarization twisting and the molecular rotation should be comparable, otherwise, the spatial symmetry about the polarization direction prevails, leading to zero orientation.
In our qualitative discussions we did not consider the initial angular velocities, but the conclusions remain valid even when the molecules have finite initial angular velocities.

A similar effect of enantioselective orientation can be achieved using THz pulses with twisted polarization. However, since the THz field couples to the molecular permanent dipole moment, unlike the case of laser pulses, two delayed orthogonally polarized THz pulses can also induce the enantioselective orientation. \cite{Tutunnikov2021Enantioselective}.

\section{Conclusions}

In this work, we have investigated the enantioselective orientation of chiral molecules excited by a shaped picosecond laser pulse and a delayed femtosecond pulse. We show that orientation is induced along the laser propagation direction, and the sign of orientation is opposite for the two enantiomers. This is similar to previous works on enantioselective orientation induced by laser pulses with twisted polarization  \cite{Yachmenev2016Detecting,Gershnabel2018Orienting,Tutunnikov2018Selective,Milner2019Controlled,Tutunnikov2019Laser,Tutunnikov2020Observation}. The use of a relatively weak, truncated picosecond pulse to induce alignment may be compared to the use of a strong femtosecond laser pulse because it avoids molecular ionization while still allowing higher degree of alignment.
Another method for achieving a high degree of alignment for a given pulse energy while minimizing molecular ionization is to use a pulse sequence consisting of multiple low-intensity pulses \cite{Averbukh2001,Averbukh2003,Averbukh2004,Lee2004,Bisgaard2004,Pinkham2007}. This approach allows the distribution of the energy, which would otherwise be delivered in a single intense pulse, over multiple low-intensity pulses. However, for the optimal degree of alignment, optimization of the sequence parameters is required.
Despite the relatively low intensities of the exciting laser pulses, a strong enantioselective orientation (the degree of orientation $\approx 0.23$) was demonstrated.
The qualitative features of the effect, including the vanishing orientation in the plane of laser polarizations, the chirality dependence of the enantioselective orientation along the laser propagation direction, and the roles of the two delayed cross-polarized laser pulses, may be understood in terms of the classical model.
The shaped picosecond laser pulse prepares a highly aligned molecular ensemble, which is followed by a cross-polarized femtosecond laser pulse. This induces asymmetric torques (depending on the off-diagonal polarizability elements which exist in chiral molecules), resulting in the nonzero orientation along the direction perpendicular to the polarizations of the two exciting pulses. During this process, the orientation along any direction in the plane of pulses' polarizations remains zero.
The enantioselective orientation may be useful for the enantiomeric excess analysis and potentially for separating mixtures of enantiomers using inhomogeneous electromagnetic fields \cite{Yachmenev2019Field}.

\begin{acknowledgments}
The author appreciates many useful discussions with Ilia Tutunnikov, Yehiam Prior, and Ilya Sh. Averbukh.
This work was supported by the Israel Science Foundation (Grant No. 746/15).
\end{acknowledgments}

\appendix

\section{Time-dependent dipole signal as a function of \texorpdfstring{$\chi_0$}{}}

\begin{figure*}[!tp]
\centering{}
\includegraphics[width=\linewidth]{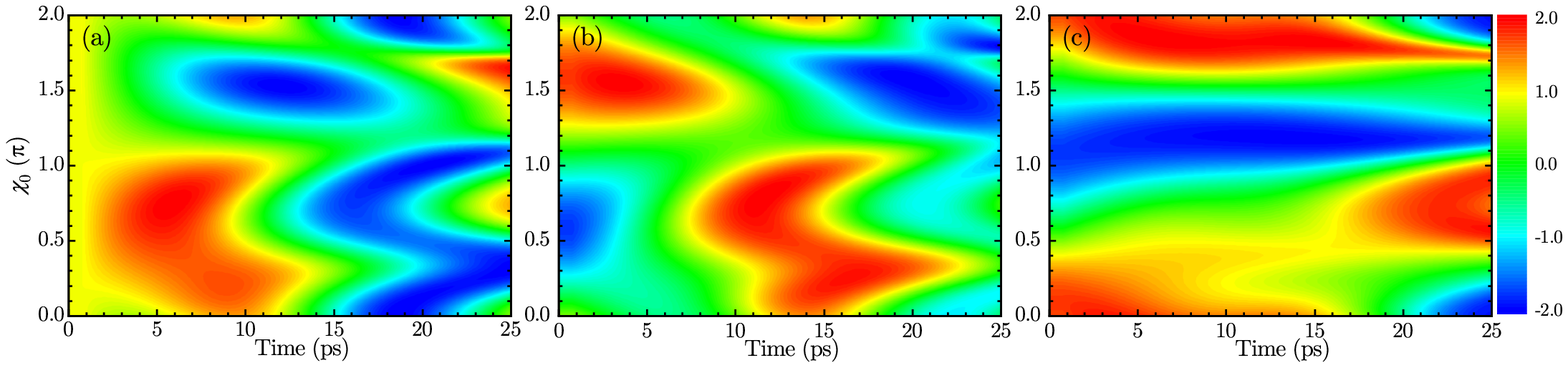}
\includegraphics[width=\linewidth]{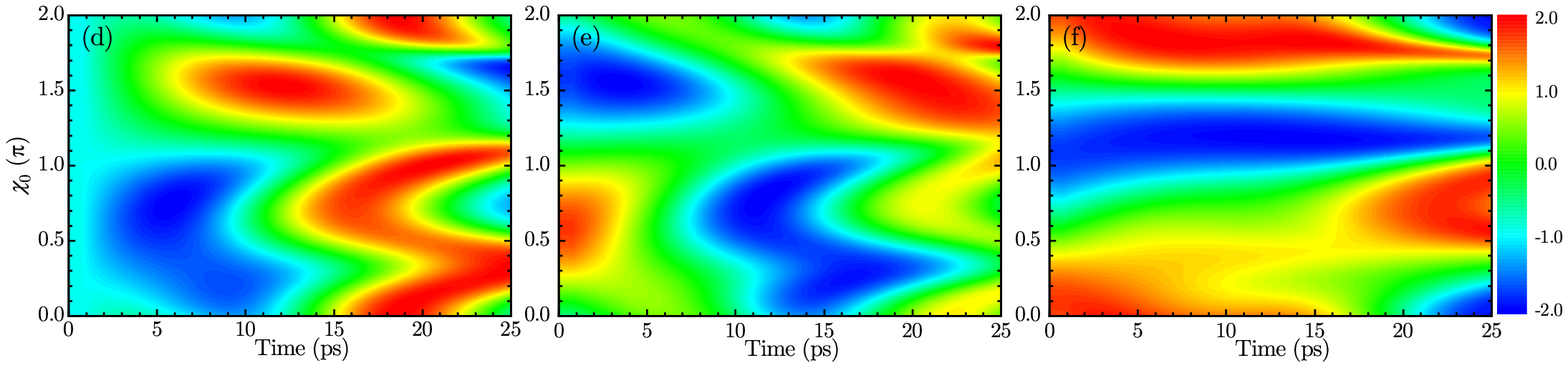}
\caption{(a, d) $X$, (b, e) $Y$, and (c, f) $Z$ projections of the dipole signal as functions of $\chi_0$ and time for the case of (\emph{R})-PPO molecules shown in Fig. \ref{fig:Classical_a}.
(a)-(c) $\phi_0=0,\,\theta_0=\pi/2$. (d)-(f) $\phi_0=\pi,\,\theta_0=\pi/2$.
The color scales are in units of Debye.
 \label{fig:distribution-RPPO}}
\end{figure*}

\begin{figure*}[!t]
\centering{}
\includegraphics[width=\linewidth]{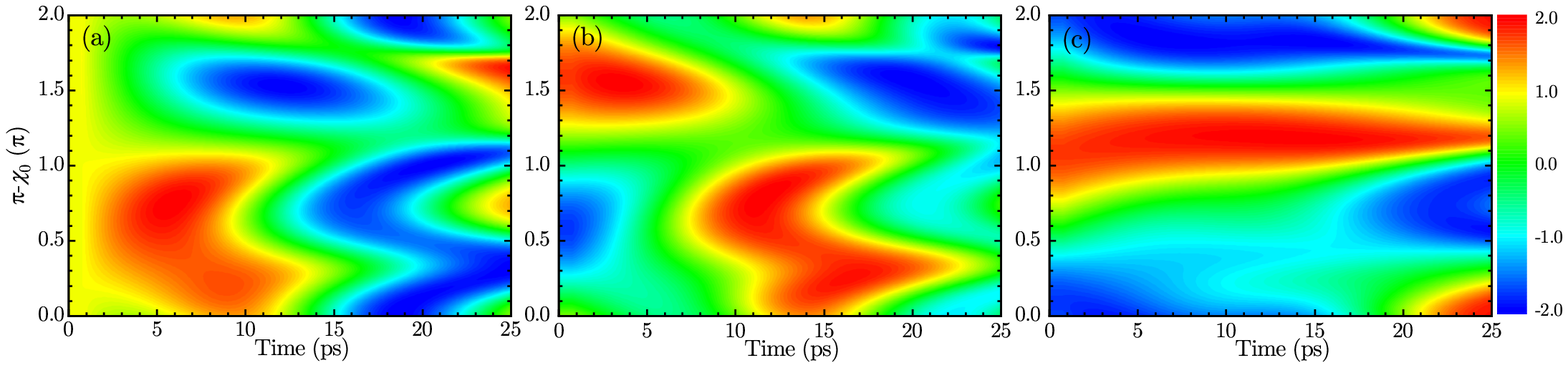}
\includegraphics[width=\linewidth]{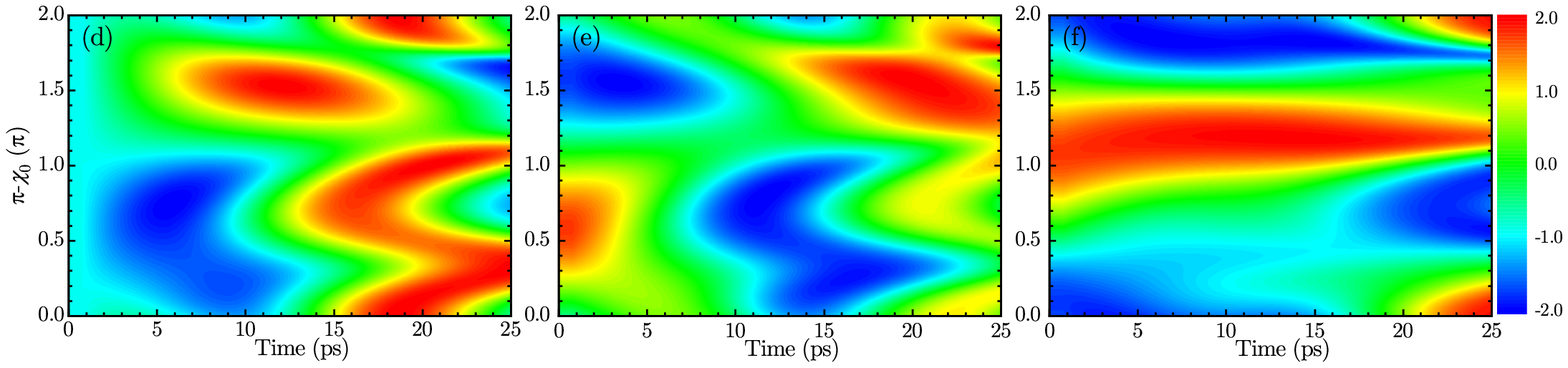}
\caption{(a, d) $X$, (b, e) $Y$, and (c, f) $Z$ projections of the dipole signal as functions of $\pi-\chi_0$ and time for the case of (\emph{S})-PPO molecules shown in Fig. \ref{fig:Classical_a}.
(a)-(c) $\phi_0=0,\,\theta_0=\pi/2$. (d)-(f) $\phi_0=\pi,\,\theta_0=\pi/2$.
The color scales are in units of Debye.
 \label{fig:distribution-SPPO}}
\end{figure*}

\begin{figure}[!t]
\centering{}
\includegraphics[width=8cm]{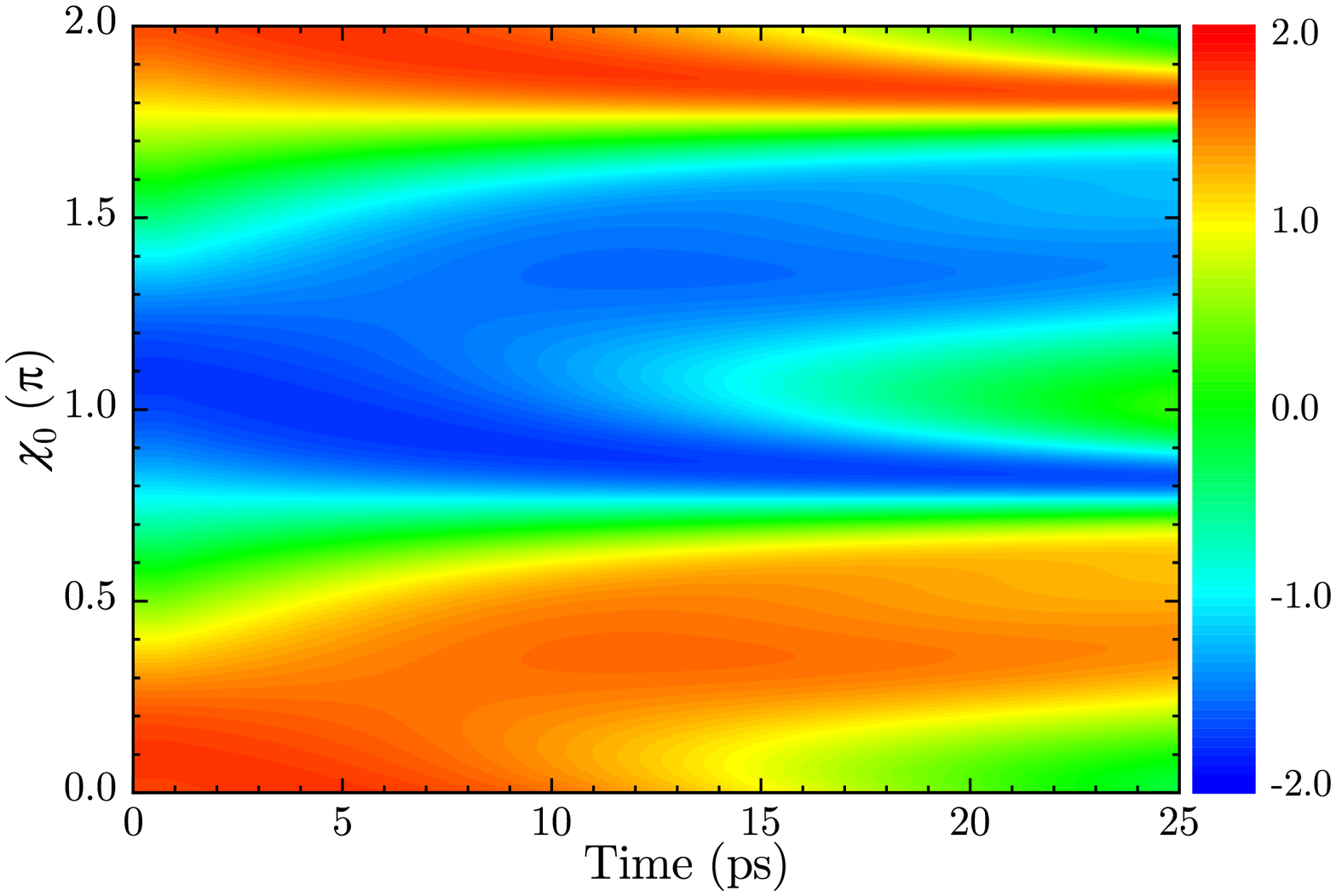}
\caption{$Z$ projection of the dipole signal as a function of $\chi_0$ and time for (\emph{R})-PPO molecules.
The color scales are in units of Debye. The initial conditions and the field parameters of the femtosecond laser pulse are the same as the case shown in Fig. \ref{fig:Classical_a} except that the femtosecond pulse is polarized along the $Y$ direction.
Note that the results for $\phi_0 = 0$ and $\phi_0=\pi$ are the same.
 \label{fig:distribution-RPPO-Ypolarized}}
\end{figure}

Figure \ref{fig:distribution-RPPO} depicts the time-dependent dipole signal as a function of $\chi_0$ for the case of (\emph{R})-PPO shown in Fig. \ref{fig:Classical_a}.
As can be seen, the dipole signals along the laboratory $X$ and $Y$ axes have opposite signs for $A(t_0)=(0, \pi/2, \chi_0)$ [panels (a) and (b)] and $B(t_0)=(\pi, \pi/2, \chi_0)$ [panels (d) and (e)].
Consequently, the ensemble-averaged dipole signals $\braket{\mu_X}$ and $\braket{\mu_Y}$ are zero.
On the other hand, Figs. \ref{fig:distribution-RPPO}(c) and \ref{fig:distribution-RPPO}(f) show that for $A(t)$ and $B(t)$, the $Z$ projections of the dipole signal are the same.
These results are consistent with the analysis presented in Sec. \ref{sec:XYdirections}.

Figure \ref{fig:distribution-SPPO} shows the case of (\emph{S})-PPO molecules.
The dipole signal of (\emph{S})-PPO along the $X$ and $Y$ directions are the same as that of (\emph{R})-PPO [$\mu_X^{(S)}(B_1)=\mu_X^{(R)}(B)$ and $\mu_Y^{(S)}(B_1)=\mu_Y^{(R)}(B)$], while $\mu_Z^{(S)}(B_1)=-\mu_Z^{(R)}(B)$.
Here $B=(\pi, \pi/2, \chi_0)$ and $B_1=(\pi, \pi/2, \pi-\chi_0)$.
According to Eq. \eqref{eq:chiral}, $\mu_j^{(S)}(B_1)=-\mu_j^{(R)}(A)$, $j=X,Y,Z$.
In addition, $\mu_j(B)=(-1)^{\delta_{jZ}+1}\mu_j(A)$ (see Sec. \ref{sec:XYdirections}), such that $\mu_j^{(S)}(B_1)=(-1)^{\delta_{jZ}}\mu_j^{(R)}(B)$.
The numerical results are consistent with the analysis, and $\mu_Z^{(S)}(B_1)=-\mu_Z^{(R)}(B)$ implies the enantioselective orientation.

Taking (\emph{R})-PPO molecule as an example, we plot $\mu_Z(t)$ as a function of $\chi_0$ in Fig. \ref{fig:distribution-RPPO-Ypolarized}.
Initially, $\phi_0 =0, \pi$ and $\theta_0=\pi/2$, corresponding to the molecular $a$ axis pointing along or against the laboratory $X$ axis. The initial temperature is $0\,\mathrm{K}$. The femtosecond laser pulse is polarized along the laboratory $Y$ axis.
As shown in Fig. \ref{fig:distribution-RPPO-Ypolarized}, $\mu_Z(\pi+\chi_0)=-\mu_Z(\chi_0)$, in agreement with the discussion in Sec. \ref{sec:roles}.
Consequently, in this case, the ensemble-averaged dipole signal along the $Z$ axis, $\braket{\mu_Z}$ vanishes.

\bibliography{references}

\begin{thebibliography}{65}%
\makeatletter
\providecommand \@ifxundefined [1]{%
 \@ifx{#1\undefined}
}%
\providecommand \@ifnum [1]{%
 \ifnum #1\expandafter \@firstoftwo
 \else \expandafter \@secondoftwo
 \fi
}%
\providecommand \@ifx [1]{%
 \ifx #1\expandafter \@firstoftwo
 \else \expandafter \@secondoftwo
 \fi
}%
\providecommand \natexlab [1]{#1}%
\providecommand \enquote  [1]{``#1''}%
\providecommand \bibnamefont  [1]{#1}%
\providecommand \bibfnamefont [1]{#1}%
\providecommand \citenamefont [1]{#1}%
\providecommand \href@noop [0]{\@secondoftwo}%
\providecommand \href [0]{\begingroup \@sanitize@url \@href}%
\providecommand \@href[1]{\@@startlink{#1}\@@href}%
\providecommand \@@href[1]{\endgroup#1\@@endlink}%
\providecommand \@sanitize@url [0]{\catcode `\\12\catcode `\$12\catcode
  `\&12\catcode `\#12\catcode `\^12\catcode `\_12\catcode `\%12\relax}%
\providecommand \@@startlink[1]{}%
\providecommand \@@endlink[0]{}%
\providecommand \url  [0]{\begingroup\@sanitize@url \@url }%
\providecommand \@url [1]{\endgroup\@href {#1}{\urlprefix }}%
\providecommand \urlprefix  [0]{URL }%
\providecommand \Eprint [0]{\href }%
\providecommand \doibase [0]{https://doi.org/}%
\providecommand \selectlanguage [0]{\@gobble}%
\providecommand \bibinfo  [0]{\@secondoftwo}%
\providecommand \bibfield  [0]{\@secondoftwo}%
\providecommand \translation [1]{[#1]}%
\providecommand \BibitemOpen [0]{}%
\providecommand \bibitemStop [0]{}%
\providecommand \bibitemNoStop [0]{.\EOS\space}%
\providecommand \EOS [0]{\spacefactor3000\relax}%
\providecommand \BibitemShut  [1]{\csname bibitem#1\endcsname}%
\let\auto@bib@innerbib\@empty
\bibitem [{\citenamefont {Cotton}(1990)}]{Cotton1990Chemical}%
  \BibitemOpen
  \bibfield  {author} {\bibinfo {author} {\bibfnamefont {F.~A.}\ \bibnamefont
  {Cotton}},\ }\href {https://www.wiley.com/en-us/Chemical Applications of
  Group Theory, 3rd Edition-p-9780471510949} {\emph {\bibinfo {title}
  {{Chemical Applications of Group Theory}}}},\ \bibinfo {edition} {3rd}\ ed.\
  (\bibinfo  {publisher} {John Wiley \& Sons, New York},\ \bibinfo {year}
  {1990})\BibitemShut {NoStop}%
\bibitem [{\citenamefont {Pasteur}(1848)}]{Pasteur1848}%
  \BibitemOpen
  \bibfield  {author} {\bibinfo {author} {\bibfnamefont {L.}~\bibnamefont
  {Pasteur}},\ }\bibfield  {title} {\bibinfo {title} {Sur les relations qui
  peuvent exister entre la forme crystalline, la composition chimique et le
  sens de la polarization rotatoire},\ }\href
  {https://wellcomecollection.org/works/rdepqfzw} {\bibfield  {journal}
  {\bibinfo  {journal} {Ann. Phys. Chem.}\ }\textbf {\bibinfo {volume} {24}},\
  \bibinfo {pages} {442} (\bibinfo {year} {1848})}\BibitemShut {NoStop}%
\bibitem [{\citenamefont {Ritchie}(1976)}]{Ritchie1976Theory}%
  \BibitemOpen
  \bibfield  {author} {\bibinfo {author} {\bibfnamefont {B.}~\bibnamefont
  {Ritchie}},\ }\bibfield  {title} {\bibinfo {title} {Theory of the angular
  distribution of photoelectrons ejected from optically active molecules and
  molecular negative ions},\ }\href {https://doi.org/10.1103/PhysRevA.13.1411}
  {\bibfield  {journal} {\bibinfo  {journal} {Phys. Rev. A}\ }\textbf {\bibinfo
  {volume} {13}},\ \bibinfo {pages} {1411} (\bibinfo {year}
  {1976})}\BibitemShut {NoStop}%
\bibitem [{\citenamefont {B\"owering}\ \emph {et~al.}(2001)\citenamefont
  {B\"owering}, \citenamefont {Lischke}, \citenamefont {Schmidtke},
  \citenamefont {M\"uller}, \citenamefont {Khalil},\ and\ \citenamefont
  {Heinzmann}}]{Bowering2001Asymmetry}%
  \BibitemOpen
  \bibfield  {author} {\bibinfo {author} {\bibfnamefont {N.}~\bibnamefont
  {B\"owering}}, \bibinfo {author} {\bibfnamefont {T.}~\bibnamefont {Lischke}},
  \bibinfo {author} {\bibfnamefont {B.}~\bibnamefont {Schmidtke}}, \bibinfo
  {author} {\bibfnamefont {N.}~\bibnamefont {M\"uller}}, \bibinfo {author}
  {\bibfnamefont {T.}~\bibnamefont {Khalil}},\ and\ \bibinfo {author}
  {\bibfnamefont {U.}~\bibnamefont {Heinzmann}},\ }\bibfield  {title} {\bibinfo
  {title} {{Asymmetry in Photoelectron Emission from Chiral Molecules Induced
  by Circularly Polarized Light}},\ }\href
  {https://doi.org/10.1103/PhysRevLett.86.1187} {\bibfield  {journal} {\bibinfo
   {journal} {Phys. Rev. Lett.}\ }\textbf {\bibinfo {volume} {86}},\ \bibinfo
  {pages} {1187} (\bibinfo {year} {2001})}\BibitemShut {NoStop}%
\bibitem [{\citenamefont {Lux}\ \emph {et~al.}(2012)\citenamefont {Lux},
  \citenamefont {Wollenhaupt}, \citenamefont {Bolze}, \citenamefont {Liang},
  \citenamefont {K{\"o}hler}, \citenamefont {Sarpe},\ and\ \citenamefont
  {Baumert}}]{Lux2011Circular}%
  \BibitemOpen
  \bibfield  {author} {\bibinfo {author} {\bibfnamefont {C.}~\bibnamefont
  {Lux}}, \bibinfo {author} {\bibfnamefont {M.}~\bibnamefont {Wollenhaupt}},
  \bibinfo {author} {\bibfnamefont {T.}~\bibnamefont {Bolze}}, \bibinfo
  {author} {\bibfnamefont {Q.}~\bibnamefont {Liang}}, \bibinfo {author}
  {\bibfnamefont {J.}~\bibnamefont {K{\"o}hler}}, \bibinfo {author}
  {\bibfnamefont {C.}~\bibnamefont {Sarpe}},\ and\ \bibinfo {author}
  {\bibfnamefont {T.}~\bibnamefont {Baumert}},\ }\bibfield  {title} {\bibinfo
  {title} {{Circular Dichroism in the Photoelectron Angular Distributions of
  Camphor and Fenchone from Multiphoton Ionization with Femtosecond Laser
  Pulses}},\ }\href {https://doi.org/10.1002/anie.201109035} {\bibfield
  {journal} {\bibinfo  {journal} {Angew. Chem. Int. Ed. Engl.}\ }\textbf
  {\bibinfo {volume} {51}},\ \bibinfo {pages} {5001} (\bibinfo {year}
  {2012})}\BibitemShut {NoStop}%
\bibitem [{\citenamefont {Beaulieu}\ \emph {et~al.}(2017)\citenamefont
  {Beaulieu}, \citenamefont {Comby}, \citenamefont {Clergerie}, \citenamefont
  {Caillat}, \citenamefont {Descamps}, \citenamefont {Dudovich}, \citenamefont
  {Fabre}, \citenamefont {G{\'e}neaux}, \citenamefont {L{\'e}gar{\'e}},
  \citenamefont {Petit}, \citenamefont {Pons}, \citenamefont {Porat},
  \citenamefont {Ruchon}, \citenamefont {Ta{\"\i}eb}, \citenamefont
  {Blanchet},\ and\ \citenamefont {Mairesse}}]{Beaulieu2017Attosecond}%
  \BibitemOpen
  \bibfield  {author} {\bibinfo {author} {\bibfnamefont {S.}~\bibnamefont
  {Beaulieu}}, \bibinfo {author} {\bibfnamefont {A.}~\bibnamefont {Comby}},
  \bibinfo {author} {\bibfnamefont {A.}~\bibnamefont {Clergerie}}, \bibinfo
  {author} {\bibfnamefont {J.}~\bibnamefont {Caillat}}, \bibinfo {author}
  {\bibfnamefont {D.}~\bibnamefont {Descamps}}, \bibinfo {author}
  {\bibfnamefont {N.}~\bibnamefont {Dudovich}}, \bibinfo {author}
  {\bibfnamefont {B.}~\bibnamefont {Fabre}}, \bibinfo {author} {\bibfnamefont
  {R.}~\bibnamefont {G{\'e}neaux}}, \bibinfo {author} {\bibfnamefont
  {F.}~\bibnamefont {L{\'e}gar{\'e}}}, \bibinfo {author} {\bibfnamefont
  {S.}~\bibnamefont {Petit}}, \bibinfo {author} {\bibfnamefont
  {B.}~\bibnamefont {Pons}}, \bibinfo {author} {\bibfnamefont {G.}~\bibnamefont
  {Porat}}, \bibinfo {author} {\bibfnamefont {T.}~\bibnamefont {Ruchon}},
  \bibinfo {author} {\bibfnamefont {R.}~\bibnamefont {Ta{\"\i}eb}}, \bibinfo
  {author} {\bibfnamefont {V.}~\bibnamefont {Blanchet}},\ and\ \bibinfo
  {author} {\bibfnamefont {Y.}~\bibnamefont {Mairesse}},\ }\bibfield  {title}
  {\bibinfo {title} {Attosecond-resolved photoionization of chiral molecules},\
  }\href {https://doi.org/10.1126/science.aao5624} {\bibfield  {journal}
  {\bibinfo  {journal} {Science}\ }\textbf {\bibinfo {volume} {358}},\ \bibinfo
  {pages} {1288} (\bibinfo {year} {2017})}\BibitemShut {NoStop}%
\bibitem [{\citenamefont {Beaulieu}\ \emph {et~al.}(2018)\citenamefont
  {Beaulieu}, \citenamefont {Comby}, \citenamefont {Descamps}, \citenamefont
  {Fabre}, \citenamefont {Garcia}, \citenamefont {G{\'e}neaux}, \citenamefont
  {Harvey}, \citenamefont {L{\'e}gar{\'e}}, \citenamefont {Ma\v{s}{\'i}n},
  \citenamefont {Nahon}, \citenamefont {Ordonez}, \citenamefont {Petit},
  \citenamefont {Pons}, \citenamefont {Mairesse}, \citenamefont {Smirnova},\
  and\ \citenamefont {Blanchet}}]{Beaulieu2018Photoexcitation}%
  \BibitemOpen
  \bibfield  {author} {\bibinfo {author} {\bibfnamefont {S.}~\bibnamefont
  {Beaulieu}}, \bibinfo {author} {\bibfnamefont {A.}~\bibnamefont {Comby}},
  \bibinfo {author} {\bibfnamefont {D.}~\bibnamefont {Descamps}}, \bibinfo
  {author} {\bibfnamefont {B.}~\bibnamefont {Fabre}}, \bibinfo {author}
  {\bibfnamefont {G.~A.}\ \bibnamefont {Garcia}}, \bibinfo {author}
  {\bibfnamefont {R.}~\bibnamefont {G{\'e}neaux}}, \bibinfo {author}
  {\bibfnamefont {A.~G.}\ \bibnamefont {Harvey}}, \bibinfo {author}
  {\bibfnamefont {F.}~\bibnamefont {L{\'e}gar{\'e}}}, \bibinfo {author}
  {\bibfnamefont {Z.}~\bibnamefont {Ma\v{s}{\'i}n}}, \bibinfo {author}
  {\bibfnamefont {L.}~\bibnamefont {Nahon}}, \bibinfo {author} {\bibfnamefont
  {A.~F.}\ \bibnamefont {Ordonez}}, \bibinfo {author} {\bibfnamefont
  {S.}~\bibnamefont {Petit}}, \bibinfo {author} {\bibfnamefont
  {B.}~\bibnamefont {Pons}}, \bibinfo {author} {\bibfnamefont {Y.}~\bibnamefont
  {Mairesse}}, \bibinfo {author} {\bibfnamefont {O.}~\bibnamefont {Smirnova}},\
  and\ \bibinfo {author} {\bibfnamefont {V.}~\bibnamefont {Blanchet}},\
  }\bibfield  {title} {\bibinfo {title} {Photoexcitation circular dichroism in
  chiral molecules},\ }\href {https://doi.org/10.1038/s41567-017-0038-z}
  {\bibfield  {journal} {\bibinfo  {journal} {Nat. Phys.}\ }\textbf {\bibinfo
  {volume} {14}},\ \bibinfo {pages} {484} (\bibinfo {year} {2018})}\BibitemShut
  {NoStop}%
\bibitem [{\citenamefont {Patterson}\ \emph {et~al.}(2013)\citenamefont
  {Patterson}, \citenamefont {Schnell},\ and\ \citenamefont
  {Doyle}}]{patterson2013enantiomer}%
  \BibitemOpen
  \bibfield  {author} {\bibinfo {author} {\bibfnamefont {D.}~\bibnamefont
  {Patterson}}, \bibinfo {author} {\bibfnamefont {M.}~\bibnamefont {Schnell}},\
  and\ \bibinfo {author} {\bibfnamefont {J.~M.}\ \bibnamefont {Doyle}},\
  }\bibfield  {title} {\bibinfo {title} {Enantiomer-specific detection of
  chiral molecules via microwave spectroscopy},\ }\href
  {https://doi.org/10.1038/nature12150} {\bibfield  {journal} {\bibinfo
  {journal} {Nature}\ }\textbf {\bibinfo {volume} {497}},\ \bibinfo {pages}
  {475} (\bibinfo {year} {2013})}\BibitemShut {NoStop}%
\bibitem [{\citenamefont {Patterson}\ and\ \citenamefont
  {Doyle}(2013)}]{Patterson2013Sensitive}%
  \BibitemOpen
  \bibfield  {author} {\bibinfo {author} {\bibfnamefont {D.}~\bibnamefont
  {Patterson}}\ and\ \bibinfo {author} {\bibfnamefont {J.~M.}\ \bibnamefont
  {Doyle}},\ }\bibfield  {title} {\bibinfo {title} {{Sensitive Chiral Analysis
  via Microwave Three-Wave Mixing}},\ }\href
  {https://doi.org/10.1103/PhysRevLett.111.023008} {\bibfield  {journal}
  {\bibinfo  {journal} {Phys. Rev. Lett.}\ }\textbf {\bibinfo {volume} {111}},\
  \bibinfo {pages} {023008} (\bibinfo {year} {2013})}\BibitemShut {NoStop}%
\bibitem [{\citenamefont {Patterson}\ and\ \citenamefont
  {Schnell}(2014)}]{Patterson2014New}%
  \BibitemOpen
  \bibfield  {author} {\bibinfo {author} {\bibfnamefont {D.}~\bibnamefont
  {Patterson}}\ and\ \bibinfo {author} {\bibfnamefont {M.}~\bibnamefont
  {Schnell}},\ }\bibfield  {title} {\bibinfo {title} {New studies on molecular
  chirality in the gas phase: enantiomer differentiation and determination of
  enantiomeric excess},\ }\href {https://doi.org/10.1039/C4CP00417E} {\bibfield
   {journal} {\bibinfo  {journal} {Phys. Chem. Chem. Phys.}\ }\textbf {\bibinfo
  {volume} {16}},\ \bibinfo {pages} {11114} (\bibinfo {year}
  {2014})}\BibitemShut {NoStop}%
\bibitem [{\citenamefont {Shubert}\ \emph {et~al.}(2014)\citenamefont
  {Shubert}, \citenamefont {Schmitz},\ and\ \citenamefont
  {Schnell}}]{Alvin2014Enantiomer}%
  \BibitemOpen
  \bibfield  {author} {\bibinfo {author} {\bibfnamefont {V.~A.}\ \bibnamefont
  {Shubert}}, \bibinfo {author} {\bibfnamefont {D.}~\bibnamefont {Schmitz}},\
  and\ \bibinfo {author} {\bibfnamefont {M.}~\bibnamefont {Schnell}},\
  }\bibfield  {title} {\bibinfo {title} {Enantiomer-sensitive spectroscopy and
  mixture analysis of chiral molecules containing two stereogenic centers --
  microwave three-wave mixing of menthone},\ }\href
  {https://doi.org/https://doi.org/10.1016/j.jms.2014.04.002} {\bibfield
  {journal} {\bibinfo  {journal} {J. Mol. Spectrosc.}\ }\textbf {\bibinfo
  {volume} {300}},\ \bibinfo {pages} {31 } (\bibinfo {year}
  {2014})}\BibitemShut {NoStop}%
\bibitem [{\citenamefont {Lehmann}(2018)}]{lehmann2018theory}%
  \BibitemOpen
  \bibfield  {author} {\bibinfo {author} {\bibfnamefont {K.~K.}\ \bibnamefont
  {Lehmann}},\ }\bibfield  {title} {\bibinfo {title} {{Theory of
  Enantiomer-Specific Microwave Spectroscopy}},\ }in\ \href
  {https://doi.org/https://doi.org/10.1016/B978-0-12-811220-5.00022-8} {\emph
  {\bibinfo {booktitle} {Frontiers and Advances in Molecular Spectroscopy}}}\
  (\bibinfo  {publisher} {Elsevier},\ \bibinfo {year} {2018})\ pp.\ \bibinfo
  {pages} {713--743}\BibitemShut {NoStop}%
\bibitem [{\citenamefont {Ye}\ \emph {et~al.}(2018)\citenamefont {Ye},
  \citenamefont {Zhang},\ and\ \citenamefont {Li}}]{Ye2018Real}%
  \BibitemOpen
  \bibfield  {author} {\bibinfo {author} {\bibfnamefont {C.}~\bibnamefont
  {Ye}}, \bibinfo {author} {\bibfnamefont {Q.}~\bibnamefont {Zhang}},\ and\
  \bibinfo {author} {\bibfnamefont {Y.}~\bibnamefont {Li}},\ }\bibfield
  {title} {\bibinfo {title} {Real single-loop cyclic three-level configuration
  of chiral molecules},\ }\href {https://doi.org/10.1103/PhysRevA.98.063401}
  {\bibfield  {journal} {\bibinfo  {journal} {Phys. Rev. A}\ }\textbf {\bibinfo
  {volume} {98}},\ \bibinfo {pages} {063401} (\bibinfo {year}
  {2018})}\BibitemShut {NoStop}%
\bibitem [{\citenamefont {Ye}\ \emph {et~al.}(2019)\citenamefont {Ye},
  \citenamefont {Zhang}, \citenamefont {Chen},\ and\ \citenamefont
  {Li}}]{Ye2019Determination}%
  \BibitemOpen
  \bibfield  {author} {\bibinfo {author} {\bibfnamefont {C.}~\bibnamefont
  {Ye}}, \bibinfo {author} {\bibfnamefont {Q.}~\bibnamefont {Zhang}}, \bibinfo
  {author} {\bibfnamefont {Y.-Y.}\ \bibnamefont {Chen}},\ and\ \bibinfo
  {author} {\bibfnamefont {Y.}~\bibnamefont {Li}},\ }\bibfield  {title}
  {\bibinfo {title} {Determination of enantiomeric excess with
  chirality-dependent ac stark effects in cyclic three-level models},\ }\href
  {https://doi.org/10.1103/PhysRevA.100.033411} {\bibfield  {journal} {\bibinfo
   {journal} {Phys. Rev. A}\ }\textbf {\bibinfo {volume} {100}},\ \bibinfo
  {pages} {033411} (\bibinfo {year} {2019})}\BibitemShut {NoStop}%
\bibitem [{\citenamefont {Leibscher}\ \emph {et~al.}(2019)\citenamefont
  {Leibscher}, \citenamefont {Giesen},\ and\ \citenamefont
  {Koch}}]{leibscher2019principles}%
  \BibitemOpen
  \bibfield  {author} {\bibinfo {author} {\bibfnamefont {M.}~\bibnamefont
  {Leibscher}}, \bibinfo {author} {\bibfnamefont {T.~F.}\ \bibnamefont
  {Giesen}},\ and\ \bibinfo {author} {\bibfnamefont {C.~P.}\ \bibnamefont
  {Koch}},\ }\bibfield  {title} {\bibinfo {title} {Principles of
  enantio-selective excitation in three-wave mixing spectroscopy of chiral
  molecules},\ }\href {https://doi.org/10.1063/1.5097406} {\bibfield  {journal}
  {\bibinfo  {journal} {J. Chem. Phys.}\ }\textbf {\bibinfo {volume} {151}},\
  \bibinfo {pages} {014302} (\bibinfo {year} {2019})}\BibitemShut {NoStop}%
\bibitem [{\citenamefont {Pitzer}\ \emph {et~al.}(2013)\citenamefont {Pitzer},
  \citenamefont {Kunitski}, \citenamefont {Johnson}, \citenamefont {Jahnke},
  \citenamefont {Sann}, \citenamefont {Sturm}, \citenamefont {{Ph. H.
  Schmidt}}, \citenamefont {Schmidt-B{\"o}cking}, \citenamefont {D{\"o}rner},
  \citenamefont {Stohner}, \citenamefont {Kiedrowski}, \citenamefont
  {Reggelin}, \citenamefont {Marquardt}, \citenamefont {Schie{\ss}er},
  \citenamefont {Berger},\ and\ \citenamefont
  {Sch{\"o}ffler}}]{Pitzer2013Direct}%
  \BibitemOpen
  \bibfield  {author} {\bibinfo {author} {\bibfnamefont {M.}~\bibnamefont
  {Pitzer}}, \bibinfo {author} {\bibfnamefont {M.}~\bibnamefont {Kunitski}},
  \bibinfo {author} {\bibfnamefont {A.~S.}\ \bibnamefont {Johnson}}, \bibinfo
  {author} {\bibfnamefont {T.}~\bibnamefont {Jahnke}}, \bibinfo {author}
  {\bibfnamefont {H.}~\bibnamefont {Sann}}, \bibinfo {author} {\bibfnamefont
  {F.}~\bibnamefont {Sturm}}, \bibinfo {author} {\bibfnamefont
  {L.}~\bibnamefont {{Ph. H. Schmidt}}}, \bibinfo {author} {\bibfnamefont
  {H.}~\bibnamefont {Schmidt-B{\"o}cking}}, \bibinfo {author} {\bibfnamefont
  {R.}~\bibnamefont {D{\"o}rner}}, \bibinfo {author} {\bibfnamefont
  {J.}~\bibnamefont {Stohner}}, \bibinfo {author} {\bibfnamefont
  {J.}~\bibnamefont {Kiedrowski}}, \bibinfo {author} {\bibfnamefont
  {M.}~\bibnamefont {Reggelin}}, \bibinfo {author} {\bibfnamefont
  {S.}~\bibnamefont {Marquardt}}, \bibinfo {author} {\bibfnamefont
  {A.}~\bibnamefont {Schie{\ss}er}}, \bibinfo {author} {\bibfnamefont
  {R.}~\bibnamefont {Berger}},\ and\ \bibinfo {author} {\bibfnamefont {M.~S.}\
  \bibnamefont {Sch{\"o}ffler}},\ }\bibfield  {title} {\bibinfo {title}
  {{Direct Determination of Absolute Molecular Stereochemistry in Gas Phase by
  Coulomb Explosion Imaging}},\ }\href
  {https://doi.org/10.1126/science.1240362} {\bibfield  {journal} {\bibinfo
  {journal} {Science}\ }\textbf {\bibinfo {volume} {341}},\ \bibinfo {pages}
  {1096} (\bibinfo {year} {2013})}\BibitemShut {NoStop}%
\bibitem [{\citenamefont {Herwig}\ \emph {et~al.}(2013)\citenamefont {Herwig},
  \citenamefont {Zawatzky}, \citenamefont {Grieser}, \citenamefont {Heber},
  \citenamefont {Jordon-Thaden}, \citenamefont {Krantz}, \citenamefont
  {Novotn{\'y}}, \citenamefont {Repnow}, \citenamefont {Schurig}, \citenamefont
  {Schwalm}, \citenamefont {Vager}, \citenamefont {Wolf}, \citenamefont
  {Trapp},\ and\ \citenamefont {Kreckel}}]{Herwig2013Imaging}%
  \BibitemOpen
  \bibfield  {author} {\bibinfo {author} {\bibfnamefont {P.}~\bibnamefont
  {Herwig}}, \bibinfo {author} {\bibfnamefont {K.}~\bibnamefont {Zawatzky}},
  \bibinfo {author} {\bibfnamefont {M.}~\bibnamefont {Grieser}}, \bibinfo
  {author} {\bibfnamefont {O.}~\bibnamefont {Heber}}, \bibinfo {author}
  {\bibfnamefont {B.}~\bibnamefont {Jordon-Thaden}}, \bibinfo {author}
  {\bibfnamefont {C.}~\bibnamefont {Krantz}}, \bibinfo {author} {\bibfnamefont
  {O.}~\bibnamefont {Novotn{\'y}}}, \bibinfo {author} {\bibfnamefont
  {R.}~\bibnamefont {Repnow}}, \bibinfo {author} {\bibfnamefont
  {V.}~\bibnamefont {Schurig}}, \bibinfo {author} {\bibfnamefont
  {D.}~\bibnamefont {Schwalm}}, \bibinfo {author} {\bibfnamefont
  {Z.}~\bibnamefont {Vager}}, \bibinfo {author} {\bibfnamefont
  {A.}~\bibnamefont {Wolf}}, \bibinfo {author} {\bibfnamefont {O.}~\bibnamefont
  {Trapp}},\ and\ \bibinfo {author} {\bibfnamefont {H.}~\bibnamefont
  {Kreckel}},\ }\bibfield  {title} {\bibinfo {title} {{Imaging the Absolute
  Configuration of a Chiral Epoxide in the Gas Phase}},\ }\href
  {https://doi.org/10.1126/science.1246549} {\bibfield  {journal} {\bibinfo
  {journal} {Science}\ }\textbf {\bibinfo {volume} {342}},\ \bibinfo {pages}
  {1084} (\bibinfo {year} {2013})}\BibitemShut {NoStop}%
\bibitem [{\citenamefont {Fehre}\ \emph {et~al.}(2019)\citenamefont {Fehre},
  \citenamefont {Eckart}, \citenamefont {Kunitski}, \citenamefont {Pitzer},
  \citenamefont {Zeller}, \citenamefont {Janke}, \citenamefont {Trabert},
  \citenamefont {Rist}, \citenamefont {Weller}, \citenamefont {Hartung},
  \citenamefont {{Ph. H. Schmidt}}, \citenamefont {Jahnke}, \citenamefont
  {Berger}, \citenamefont {D{\"o}rner},\ and\ \citenamefont
  {Sch{\"o}ffler}}]{Fehre2019Enantioselective}%
  \BibitemOpen
  \bibfield  {author} {\bibinfo {author} {\bibfnamefont {K.}~\bibnamefont
  {Fehre}}, \bibinfo {author} {\bibfnamefont {S.}~\bibnamefont {Eckart}},
  \bibinfo {author} {\bibfnamefont {M.}~\bibnamefont {Kunitski}}, \bibinfo
  {author} {\bibfnamefont {M.}~\bibnamefont {Pitzer}}, \bibinfo {author}
  {\bibfnamefont {S.}~\bibnamefont {Zeller}}, \bibinfo {author} {\bibfnamefont
  {C.}~\bibnamefont {Janke}}, \bibinfo {author} {\bibfnamefont
  {D.}~\bibnamefont {Trabert}}, \bibinfo {author} {\bibfnamefont
  {J.}~\bibnamefont {Rist}}, \bibinfo {author} {\bibfnamefont {M.}~\bibnamefont
  {Weller}}, \bibinfo {author} {\bibfnamefont {A.}~\bibnamefont {Hartung}},
  \bibinfo {author} {\bibfnamefont {L.}~\bibnamefont {{Ph. H. Schmidt}}},
  \bibinfo {author} {\bibfnamefont {T.}~\bibnamefont {Jahnke}}, \bibinfo
  {author} {\bibfnamefont {R.}~\bibnamefont {Berger}}, \bibinfo {author}
  {\bibfnamefont {R.}~\bibnamefont {D{\"o}rner}},\ and\ \bibinfo {author}
  {\bibfnamefont {M.~S.}\ \bibnamefont {Sch{\"o}ffler}},\ }\bibfield  {title}
  {\bibinfo {title} {Enantioselective fragmentation of an achiral molecule in a
  strong laser field},\ }\href {https://doi.org/10.1126/sciadv.aau7923}
  {\bibfield  {journal} {\bibinfo  {journal} {Sci. Adv.}\ }\textbf {\bibinfo
  {volume} {5}},\ \bibinfo {pages} {eaau7923} (\bibinfo {year}
  {2019})}\BibitemShut {NoStop}%
\bibitem [{\citenamefont {Cireasa}\ \emph {et~al.}(2015)\citenamefont
  {Cireasa}, \citenamefont {Boguslavskiy}, \citenamefont {Pons}, \citenamefont
  {Wong}, \citenamefont {Descamps}, \citenamefont {Petit}, \citenamefont {Ruf},
  \citenamefont {Thir{\'e}}, \citenamefont {Ferr{\'e}}, \citenamefont {Suarez},
  \citenamefont {Higuet}, \citenamefont {Schmidt}, \citenamefont {Alharbi},
  \citenamefont {L{\'e}gar{\'e}}, \citenamefont {Blanchet}, \citenamefont
  {Fabre}, \citenamefont {Patchkovskii}, \citenamefont {Smirnova},
  \citenamefont {Mairesse},\ and\ \citenamefont
  {Bhardwaj}}]{Cireasa2015Probing}%
  \BibitemOpen
  \bibfield  {author} {\bibinfo {author} {\bibfnamefont {R.}~\bibnamefont
  {Cireasa}}, \bibinfo {author} {\bibfnamefont {A.~E.}\ \bibnamefont
  {Boguslavskiy}}, \bibinfo {author} {\bibfnamefont {B.}~\bibnamefont {Pons}},
  \bibinfo {author} {\bibfnamefont {M.~C.~H.}\ \bibnamefont {Wong}}, \bibinfo
  {author} {\bibfnamefont {D.}~\bibnamefont {Descamps}}, \bibinfo {author}
  {\bibfnamefont {S.}~\bibnamefont {Petit}}, \bibinfo {author} {\bibfnamefont
  {H.}~\bibnamefont {Ruf}}, \bibinfo {author} {\bibfnamefont {N.}~\bibnamefont
  {Thir{\'e}}}, \bibinfo {author} {\bibfnamefont {A.}~\bibnamefont
  {Ferr{\'e}}}, \bibinfo {author} {\bibfnamefont {J.}~\bibnamefont {Suarez}},
  \bibinfo {author} {\bibfnamefont {J.}~\bibnamefont {Higuet}}, \bibinfo
  {author} {\bibfnamefont {B.~E.}\ \bibnamefont {Schmidt}}, \bibinfo {author}
  {\bibfnamefont {A.~F.}\ \bibnamefont {Alharbi}}, \bibinfo {author}
  {\bibfnamefont {F.}~\bibnamefont {L{\'e}gar{\'e}}}, \bibinfo {author}
  {\bibfnamefont {V.}~\bibnamefont {Blanchet}}, \bibinfo {author}
  {\bibfnamefont {B.}~\bibnamefont {Fabre}}, \bibinfo {author} {\bibfnamefont
  {S.}~\bibnamefont {Patchkovskii}}, \bibinfo {author} {\bibfnamefont
  {O.}~\bibnamefont {Smirnova}}, \bibinfo {author} {\bibfnamefont
  {Y.}~\bibnamefont {Mairesse}},\ and\ \bibinfo {author} {\bibfnamefont
  {V.~R.}\ \bibnamefont {Bhardwaj}},\ }\bibfield  {title} {\bibinfo {title}
  {Probing molecular chirality on a sub-femtosecond timescale},\ }\href
  {https://doi.org/10.1038/nphys3369} {\bibfield  {journal} {\bibinfo
  {journal} {Nat. Phys.}\ }\textbf {\bibinfo {volume} {11}},\ \bibinfo {pages}
  {654} (\bibinfo {year} {2015})}\BibitemShut {NoStop}%
\bibitem [{\citenamefont {Banerjee-Ghosh}\ \emph {et~al.}(2018)\citenamefont
  {Banerjee-Ghosh}, \citenamefont {Ben~Dor}, \citenamefont {Tassinari},
  \citenamefont {Capua}, \citenamefont {Yochelis}, \citenamefont {Capua},
  \citenamefont {Yang}, \citenamefont {Parkin}, \citenamefont {Sarkar},
  \citenamefont {Kronik}, \citenamefont {Baczewski}, \citenamefont {Naaman},\
  and\ \citenamefont {Paltiel}}]{Banerjee-Ghosh2018}%
  \BibitemOpen
  \bibfield  {author} {\bibinfo {author} {\bibfnamefont {K.}~\bibnamefont
  {Banerjee-Ghosh}}, \bibinfo {author} {\bibfnamefont {O.}~\bibnamefont
  {Ben~Dor}}, \bibinfo {author} {\bibfnamefont {F.}~\bibnamefont {Tassinari}},
  \bibinfo {author} {\bibfnamefont {E.}~\bibnamefont {Capua}}, \bibinfo
  {author} {\bibfnamefont {S.}~\bibnamefont {Yochelis}}, \bibinfo {author}
  {\bibfnamefont {A.}~\bibnamefont {Capua}}, \bibinfo {author} {\bibfnamefont
  {S.-H.}\ \bibnamefont {Yang}}, \bibinfo {author} {\bibfnamefont {S.~S.~P.}\
  \bibnamefont {Parkin}}, \bibinfo {author} {\bibfnamefont {S.}~\bibnamefont
  {Sarkar}}, \bibinfo {author} {\bibfnamefont {L.}~\bibnamefont {Kronik}},
  \bibinfo {author} {\bibfnamefont {L.~T.}\ \bibnamefont {Baczewski}}, \bibinfo
  {author} {\bibfnamefont {R.}~\bibnamefont {Naaman}},\ and\ \bibinfo {author}
  {\bibfnamefont {Y.}~\bibnamefont {Paltiel}},\ }\bibfield  {title} {\bibinfo
  {title} {Separation of enantiomers by their enantiospecific interaction with
  achiral magnetic substrates},\ }\href
  {https://doi.org/10.1126/science.aar4265} {\bibfield  {journal} {\bibinfo
  {journal} {Science}\ }\textbf {\bibinfo {volume} {360}},\ \bibinfo {pages}
  {1331} (\bibinfo {year} {2018})}\BibitemShut {NoStop}%
\bibitem [{\citenamefont {Yachmenev}\ and\ \citenamefont
  {Yurchenko}(2016)}]{Yachmenev2016Detecting}%
  \BibitemOpen
  \bibfield  {author} {\bibinfo {author} {\bibfnamefont {A.}~\bibnamefont
  {Yachmenev}}\ and\ \bibinfo {author} {\bibfnamefont {S.~N.}\ \bibnamefont
  {Yurchenko}},\ }\bibfield  {title} {\bibinfo {title} {{Detecting Chirality in
  Molecules by Linearly Polarized Laser Fields}},\ }\href
  {https://doi.org/10.1103/PhysRevLett.117.033001} {\bibfield  {journal}
  {\bibinfo  {journal} {Phys. Rev. Lett.}\ }\textbf {\bibinfo {volume} {117}},\
  \bibinfo {pages} {033001} (\bibinfo {year} {2016})}\BibitemShut {NoStop}%
\bibitem [{\citenamefont {Gershnabel}\ and\ \citenamefont {{Sh.
  Averbukh}}(2018)}]{Gershnabel2018Orienting}%
  \BibitemOpen
  \bibfield  {author} {\bibinfo {author} {\bibfnamefont {E.}~\bibnamefont
  {Gershnabel}}\ and\ \bibinfo {author} {\bibfnamefont {I.}~\bibnamefont {{Sh.
  Averbukh}}},\ }\bibfield  {title} {\bibinfo {title} {Orienting asymmetric
  molecules by laser fields with twisted polarization},\ }\href
  {https://doi.org/10.1103/PhysRevLett.120.083204} {\bibfield  {journal}
  {\bibinfo  {journal} {Phys. Rev. Lett.}\ }\textbf {\bibinfo {volume} {120}},\
  \bibinfo {pages} {083204} (\bibinfo {year} {2018})}\BibitemShut {NoStop}%
\bibitem [{\citenamefont {Tutunnikov}\ \emph {et~al.}(2018)\citenamefont
  {Tutunnikov}, \citenamefont {Gershnabel}, \citenamefont {Gold},\ and\
  \citenamefont {{Sh. Averbukh}}}]{Tutunnikov2018Selective}%
  \BibitemOpen
  \bibfield  {author} {\bibinfo {author} {\bibfnamefont {I.}~\bibnamefont
  {Tutunnikov}}, \bibinfo {author} {\bibfnamefont {E.}~\bibnamefont
  {Gershnabel}}, \bibinfo {author} {\bibfnamefont {S.}~\bibnamefont {Gold}},\
  and\ \bibinfo {author} {\bibfnamefont {I.}~\bibnamefont {{Sh. Averbukh}}},\
  }\bibfield  {title} {\bibinfo {title} {{Selective Orientation of Chiral
  Molecules by Laser Fields with Twisted Polarization}},\ }\href
  {https://doi.org/10.1021/acs.jpclett.7b03416} {\bibfield  {journal} {\bibinfo
   {journal} {J. Phys. Chem. Lett.}\ }\textbf {\bibinfo {volume} {9}},\
  \bibinfo {pages} {1105} (\bibinfo {year} {2018})}\BibitemShut {NoStop}%
\bibitem [{\citenamefont {Milner}\ \emph {et~al.}(2019)\citenamefont {Milner},
  \citenamefont {Fordyce}, \citenamefont {MacPhail-Bartley}, \citenamefont
  {Wasserman}, \citenamefont {Milner}, \citenamefont {Tutunnikov},\ and\
  \citenamefont {{Sh. Averbukh}}}]{Milner2019Controlled}%
  \BibitemOpen
  \bibfield  {author} {\bibinfo {author} {\bibfnamefont {A.~A.}\ \bibnamefont
  {Milner}}, \bibinfo {author} {\bibfnamefont {J.~A.~M.}\ \bibnamefont
  {Fordyce}}, \bibinfo {author} {\bibfnamefont {I.}~\bibnamefont
  {MacPhail-Bartley}}, \bibinfo {author} {\bibfnamefont {W.}~\bibnamefont
  {Wasserman}}, \bibinfo {author} {\bibfnamefont {V.}~\bibnamefont {Milner}},
  \bibinfo {author} {\bibfnamefont {I.}~\bibnamefont {Tutunnikov}},\ and\
  \bibinfo {author} {\bibfnamefont {I.}~\bibnamefont {{Sh. Averbukh}}},\
  }\bibfield  {title} {\bibinfo {title} {{Controlled Enantioselective
  Orientation of Chiral Molecules with an Optical Centrifuge}},\ }\href
  {https://doi.org/10.1103/PhysRevLett.122.223201} {\bibfield  {journal}
  {\bibinfo  {journal} {Phys. Rev. Lett.}\ }\textbf {\bibinfo {volume} {122}},\
  \bibinfo {pages} {223201} (\bibinfo {year} {2019})}\BibitemShut {NoStop}%
\bibitem [{\citenamefont {Tutunnikov}\ \emph {et~al.}(2019)\citenamefont
  {Tutunnikov}, \citenamefont {Flo\ss{}}, \citenamefont {Gershnabel},
  \citenamefont {Brumer},\ and\ \citenamefont {{Sh.
  Averbukh}}}]{Tutunnikov2019Laser}%
  \BibitemOpen
  \bibfield  {author} {\bibinfo {author} {\bibfnamefont {I.}~\bibnamefont
  {Tutunnikov}}, \bibinfo {author} {\bibfnamefont {J.}~\bibnamefont
  {Flo\ss{}}}, \bibinfo {author} {\bibfnamefont {E.}~\bibnamefont
  {Gershnabel}}, \bibinfo {author} {\bibfnamefont {P.}~\bibnamefont {Brumer}},\
  and\ \bibinfo {author} {\bibfnamefont {I.}~\bibnamefont {{Sh. Averbukh}}},\
  }\bibfield  {title} {\bibinfo {title} {Laser-induced persistent orientation
  of chiral molecules},\ }\href {https://doi.org/10.1103/PhysRevA.100.043406}
  {\bibfield  {journal} {\bibinfo  {journal} {Phys. Rev. A}\ }\textbf {\bibinfo
  {volume} {100}},\ \bibinfo {pages} {043406} (\bibinfo {year}
  {2019})}\BibitemShut {NoStop}%
\bibitem [{\citenamefont {Tutunnikov}\ \emph {et~al.}(2020)\citenamefont
  {Tutunnikov}, \citenamefont {Flo\ss{}}, \citenamefont {Gershnabel},
  \citenamefont {Brumer}, \citenamefont {{Sh. Averbukh}}, \citenamefont
  {Milner},\ and\ \citenamefont {Milner}}]{Tutunnikov2020Observation}%
  \BibitemOpen
  \bibfield  {author} {\bibinfo {author} {\bibfnamefont {I.}~\bibnamefont
  {Tutunnikov}}, \bibinfo {author} {\bibfnamefont {J.}~\bibnamefont
  {Flo\ss{}}}, \bibinfo {author} {\bibfnamefont {E.}~\bibnamefont
  {Gershnabel}}, \bibinfo {author} {\bibfnamefont {P.}~\bibnamefont {Brumer}},
  \bibinfo {author} {\bibfnamefont {I.}~\bibnamefont {{Sh. Averbukh}}},
  \bibinfo {author} {\bibfnamefont {A.~A.}\ \bibnamefont {Milner}},\ and\
  \bibinfo {author} {\bibfnamefont {V.}~\bibnamefont {Milner}},\ }\bibfield
  {title} {\bibinfo {title} {Observation of persistent orientation of chiral
  molecules by a laser field with twisted polarization},\ }\href
  {https://doi.org/10.1103/PhysRevA.101.021403} {\bibfield  {journal} {\bibinfo
   {journal} {Phys. Rev. A}\ }\textbf {\bibinfo {volume} {101}},\ \bibinfo
  {pages} {021403(R)} (\bibinfo {year} {2020})}\BibitemShut {NoStop}%
\bibitem [{\citenamefont {Fleischer}\ \emph {et~al.}(2009)\citenamefont
  {Fleischer}, \citenamefont {Khodorkovsky}, \citenamefont {Prior},\ and\
  \citenamefont {{Sh. Averbukh}}}]{Fleischer2009Controlling}%
  \BibitemOpen
  \bibfield  {author} {\bibinfo {author} {\bibfnamefont {S.}~\bibnamefont
  {Fleischer}}, \bibinfo {author} {\bibfnamefont {Y.}~\bibnamefont
  {Khodorkovsky}}, \bibinfo {author} {\bibfnamefont {Y.}~\bibnamefont
  {Prior}},\ and\ \bibinfo {author} {\bibfnamefont {I.}~\bibnamefont {{Sh.
  Averbukh}}},\ }\bibfield  {title} {\bibinfo {title} {Controlling the sense of
  molecular rotation},\ }\href {https://doi.org/10.1088/1367-2630/11/10/105039}
  {\bibfield  {journal} {\bibinfo  {journal} {New J. Phys.}\ }\textbf {\bibinfo
  {volume} {11}},\ \bibinfo {pages} {105039} (\bibinfo {year}
  {2009})}\BibitemShut {NoStop}%
\bibitem [{\citenamefont {Kitano}\ \emph {et~al.}(2009)\citenamefont {Kitano},
  \citenamefont {Hasegawa},\ and\ \citenamefont
  {Ohshima}}]{Kitano2009Ultrafast}%
  \BibitemOpen
  \bibfield  {author} {\bibinfo {author} {\bibfnamefont {K.}~\bibnamefont
  {Kitano}}, \bibinfo {author} {\bibfnamefont {H.}~\bibnamefont {Hasegawa}},\
  and\ \bibinfo {author} {\bibfnamefont {Y.}~\bibnamefont {Ohshima}},\
  }\bibfield  {title} {\bibinfo {title} {{Ultrafast Angular Momentum
  Orientation by Linearly Polarized Laser Fields}},\ }\href
  {https://doi.org/10.1103/PhysRevLett.103.223002} {\bibfield  {journal}
  {\bibinfo  {journal} {Phys. Rev. Lett.}\ }\textbf {\bibinfo {volume} {103}},\
  \bibinfo {pages} {223002} (\bibinfo {year} {2009})}\BibitemShut {NoStop}%
\bibitem [{\citenamefont {Khodorkovsky}\ \emph {et~al.}(2011)\citenamefont
  {Khodorkovsky}, \citenamefont {Kitano}, \citenamefont {Hasegawa},
  \citenamefont {Ohshima},\ and\ \citenamefont {{Sh.
  Averbukh}}}]{Khodorkovsky2011Controlling}%
  \BibitemOpen
  \bibfield  {author} {\bibinfo {author} {\bibfnamefont {Y.}~\bibnamefont
  {Khodorkovsky}}, \bibinfo {author} {\bibfnamefont {K.}~\bibnamefont
  {Kitano}}, \bibinfo {author} {\bibfnamefont {H.}~\bibnamefont {Hasegawa}},
  \bibinfo {author} {\bibfnamefont {Y.}~\bibnamefont {Ohshima}},\ and\ \bibinfo
  {author} {\bibfnamefont {I.}~\bibnamefont {{Sh. Averbukh}}},\ }\bibfield
  {title} {\bibinfo {title} {{Controlling the sense of molecular rotation:
  Classical versus quantum analysis}},\ }\href
  {https://doi.org/10.1103/PhysRevA.83.023423} {\bibfield  {journal} {\bibinfo
  {journal} {Phys. Rev. A}\ }\textbf {\bibinfo {volume} {83}},\ \bibinfo
  {pages} {023423} (\bibinfo {year} {2011})}\BibitemShut {NoStop}%
\bibitem [{\citenamefont {Karczmarek}\ \emph {et~al.}(1999)\citenamefont
  {Karczmarek}, \citenamefont {Wright}, \citenamefont {Corkum},\ and\
  \citenamefont {Ivanov}}]{Karczmarek1999Optical}%
  \BibitemOpen
  \bibfield  {author} {\bibinfo {author} {\bibfnamefont {J.}~\bibnamefont
  {Karczmarek}}, \bibinfo {author} {\bibfnamefont {J.}~\bibnamefont {Wright}},
  \bibinfo {author} {\bibfnamefont {P.}~\bibnamefont {Corkum}},\ and\ \bibinfo
  {author} {\bibfnamefont {M.}~\bibnamefont {Ivanov}},\ }\bibfield  {title}
  {\bibinfo {title} {{Optical Centrifuge for Molecules}},\ }\href
  {https://doi.org/10.1103/PhysRevLett.82.3420} {\bibfield  {journal} {\bibinfo
   {journal} {Phys. Rev. Lett.}\ }\textbf {\bibinfo {volume} {82}},\ \bibinfo
  {pages} {3420} (\bibinfo {year} {1999})}\BibitemShut {NoStop}%
\bibitem [{\citenamefont {Villeneuve}\ \emph {et~al.}(2000)\citenamefont
  {Villeneuve}, \citenamefont {Aseyev}, \citenamefont {Dietrich}, \citenamefont
  {Spanner}, \citenamefont {{Yu. Ivanov}},\ and\ \citenamefont
  {Corkum}}]{Villeneuve2000Forced}%
  \BibitemOpen
  \bibfield  {author} {\bibinfo {author} {\bibfnamefont {D.~M.}\ \bibnamefont
  {Villeneuve}}, \bibinfo {author} {\bibfnamefont {S.~A.}\ \bibnamefont
  {Aseyev}}, \bibinfo {author} {\bibfnamefont {P.}~\bibnamefont {Dietrich}},
  \bibinfo {author} {\bibfnamefont {M.}~\bibnamefont {Spanner}}, \bibinfo
  {author} {\bibfnamefont {M.}~\bibnamefont {{Yu. Ivanov}}},\ and\ \bibinfo
  {author} {\bibfnamefont {P.~B.}\ \bibnamefont {Corkum}},\ }\bibfield  {title}
  {\bibinfo {title} {{Forced Molecular Rotation in an Optical Centrifuge}},\
  }\href {https://doi.org/10.1103/PhysRevLett.85.542} {\bibfield  {journal}
  {\bibinfo  {journal} {Phys. Rev. Lett.}\ }\textbf {\bibinfo {volume} {85}},\
  \bibinfo {pages} {542} (\bibinfo {year} {2000})}\BibitemShut {NoStop}%
\bibitem [{\citenamefont {Yuan}\ \emph {et~al.}(2011)\citenamefont {Yuan},
  \citenamefont {Teitelbaum}, \citenamefont {Robinson},\ and\ \citenamefont
  {Mullin}}]{Yuan2011Dynamics}%
  \BibitemOpen
  \bibfield  {author} {\bibinfo {author} {\bibfnamefont {L.}~\bibnamefont
  {Yuan}}, \bibinfo {author} {\bibfnamefont {S.~W.}\ \bibnamefont
  {Teitelbaum}}, \bibinfo {author} {\bibfnamefont {A.}~\bibnamefont
  {Robinson}},\ and\ \bibinfo {author} {\bibfnamefont {A.~S.}\ \bibnamefont
  {Mullin}},\ }\bibfield  {title} {\bibinfo {title} {Dynamics of molecules in
  extreme rotational states},\ }\href {https://doi.org/10.1073/pnas.1018669108}
  {\bibfield  {journal} {\bibinfo  {journal} {Proc. Natl. Acad. Sci. U.S.A.}\
  }\textbf {\bibinfo {volume} {108}},\ \bibinfo {pages} {6872} (\bibinfo {year}
  {2011})}\BibitemShut {NoStop}%
\bibitem [{\citenamefont {Korobenko}\ \emph {et~al.}(2014)\citenamefont
  {Korobenko}, \citenamefont {Milner},\ and\ \citenamefont
  {Milner}}]{Korobenko2014Direct}%
  \BibitemOpen
  \bibfield  {author} {\bibinfo {author} {\bibfnamefont {A.}~\bibnamefont
  {Korobenko}}, \bibinfo {author} {\bibfnamefont {A.~A.}\ \bibnamefont
  {Milner}},\ and\ \bibinfo {author} {\bibfnamefont {V.}~\bibnamefont
  {Milner}},\ }\bibfield  {title} {\bibinfo {title} {{Direct Observation,
  Study, and Control of Molecular Superrotors}},\ }\href
  {https://doi.org/10.1103/PhysRevLett.112.113004} {\bibfield  {journal}
  {\bibinfo  {journal} {Phys. Rev. Lett.}\ }\textbf {\bibinfo {volume} {112}},\
  \bibinfo {pages} {113004} (\bibinfo {year} {2014})}\BibitemShut {NoStop}%
\bibitem [{\citenamefont {Korobenko}(2018)}]{Korobenko2018Control}%
  \BibitemOpen
  \bibfield  {author} {\bibinfo {author} {\bibfnamefont {A.}~\bibnamefont
  {Korobenko}},\ }\bibfield  {title} {\bibinfo {title} {Control of molecular
  rotation with an optical centrifuge},\ }\href
  {https://doi.org/10.1088/1361-6455/aadd56} {\bibfield  {journal} {\bibinfo
  {journal} {J. Phys. B}\ }\textbf {\bibinfo {volume} {51}},\ \bibinfo {pages}
  {203001} (\bibinfo {year} {2018})}\BibitemShut {NoStop}%
\bibitem [{\citenamefont {Zhdanovich}\ \emph {et~al.}(2011)\citenamefont
  {Zhdanovich}, \citenamefont {Milner}, \citenamefont {Bloomquist},
  \citenamefont {Flo\ss{}}, \citenamefont {{Sh. Averbukh}}, \citenamefont
  {Hepburn},\ and\ \citenamefont {Milner}}]{Zhdanovich2011Control}%
  \BibitemOpen
  \bibfield  {author} {\bibinfo {author} {\bibfnamefont {S.}~\bibnamefont
  {Zhdanovich}}, \bibinfo {author} {\bibfnamefont {A.~A.}\ \bibnamefont
  {Milner}}, \bibinfo {author} {\bibfnamefont {C.}~\bibnamefont {Bloomquist}},
  \bibinfo {author} {\bibfnamefont {J.}~\bibnamefont {Flo\ss{}}}, \bibinfo
  {author} {\bibfnamefont {I.}~\bibnamefont {{Sh. Averbukh}}}, \bibinfo
  {author} {\bibfnamefont {J.~W.}\ \bibnamefont {Hepburn}},\ and\ \bibinfo
  {author} {\bibfnamefont {V.}~\bibnamefont {Milner}},\ }\bibfield  {title}
  {\bibinfo {title} {{Control of Molecular Rotation with a Chiral Train of
  Ultrashort Pulses}},\ }\href {https://doi.org/10.1103/PhysRevLett.107.243004}
  {\bibfield  {journal} {\bibinfo  {journal} {Phys. Rev. Lett.}\ }\textbf
  {\bibinfo {volume} {107}},\ \bibinfo {pages} {243004} (\bibinfo {year}
  {2011})}\BibitemShut {NoStop}%
\bibitem [{\citenamefont {Flo\ss{}}\ and\ \citenamefont {{Sh.
  Averbukh}}(2012)}]{Johannes2012Molecular}%
  \BibitemOpen
  \bibfield  {author} {\bibinfo {author} {\bibfnamefont {J.}~\bibnamefont
  {Flo\ss{}}}\ and\ \bibinfo {author} {\bibfnamefont {I.}~\bibnamefont {{Sh.
  Averbukh}}},\ }\bibfield  {title} {\bibinfo {title} {Molecular spinning by a
  chiral train of short laser pulses},\ }\href
  {https://doi.org/10.1103/PhysRevA.86.063414} {\bibfield  {journal} {\bibinfo
  {journal} {Phys. Rev. A}\ }\textbf {\bibinfo {volume} {86}},\ \bibinfo
  {pages} {063414} (\bibinfo {year} {2012})}\BibitemShut {NoStop}%
\bibitem [{\citenamefont {Kida}\ \emph {et~al.}(2008)\citenamefont {Kida},
  \citenamefont {Zaitsu},\ and\ \citenamefont {Imasaka}}]{Kida2008Stimulated}%
  \BibitemOpen
  \bibfield  {author} {\bibinfo {author} {\bibfnamefont {Y.}~\bibnamefont
  {Kida}}, \bibinfo {author} {\bibfnamefont {S.-i.}\ \bibnamefont {Zaitsu}},\
  and\ \bibinfo {author} {\bibfnamefont {T.}~\bibnamefont {Imasaka}},\
  }\bibfield  {title} {\bibinfo {title} {{Stimulated rotational Raman
  scattering by a polarization-modulated femtosecond pulse}},\ }\href
  {https://doi.org/10.1103/PhysRevA.77.063802} {\bibfield  {journal} {\bibinfo
  {journal} {Phys. Rev. A}\ }\textbf {\bibinfo {volume} {77}},\ \bibinfo
  {pages} {063802} (\bibinfo {year} {2008})}\BibitemShut {NoStop}%
\bibitem [{\citenamefont {Kida}\ \emph {et~al.}(2009)\citenamefont {Kida},
  \citenamefont {Zaitsu},\ and\ \citenamefont {Imasaka}}]{Kida2009Coherent}%
  \BibitemOpen
  \bibfield  {author} {\bibinfo {author} {\bibfnamefont {Y.}~\bibnamefont
  {Kida}}, \bibinfo {author} {\bibfnamefont {S.-i.}\ \bibnamefont {Zaitsu}},\
  and\ \bibinfo {author} {\bibfnamefont {T.}~\bibnamefont {Imasaka}},\
  }\bibfield  {title} {\bibinfo {title} {Coherent molecular rotations induced
  by a femtosecond pulse consisting of two orthogonally polarized pulses},\
  }\href {https://doi.org/10.1103/PhysRevA.80.021805} {\bibfield  {journal}
  {\bibinfo  {journal} {Phys. Rev. A}\ }\textbf {\bibinfo {volume} {80}},\
  \bibinfo {pages} {021805(R)} (\bibinfo {year} {2009})}\BibitemShut {NoStop}%
\bibitem [{\citenamefont {Karras}\ \emph {et~al.}(2015)\citenamefont {Karras},
  \citenamefont {Ndong}, \citenamefont {Hertz}, \citenamefont {Sugny},
  \citenamefont {Billard}, \citenamefont {Lavorel},\ and\ \citenamefont
  {Faucher}}]{Karras2015Polarization}%
  \BibitemOpen
  \bibfield  {author} {\bibinfo {author} {\bibfnamefont {G.}~\bibnamefont
  {Karras}}, \bibinfo {author} {\bibfnamefont {M.}~\bibnamefont {Ndong}},
  \bibinfo {author} {\bibfnamefont {E.}~\bibnamefont {Hertz}}, \bibinfo
  {author} {\bibfnamefont {D.}~\bibnamefont {Sugny}}, \bibinfo {author}
  {\bibfnamefont {F.}~\bibnamefont {Billard}}, \bibinfo {author} {\bibfnamefont
  {B.}~\bibnamefont {Lavorel}},\ and\ \bibinfo {author} {\bibfnamefont
  {O.}~\bibnamefont {Faucher}},\ }\bibfield  {title} {\bibinfo {title}
  {{Polarization Shaping for Unidirectional Rotational Motion of Molecules}},\
  }\href {https://doi.org/10.1103/PhysRevLett.114.103001} {\bibfield  {journal}
  {\bibinfo  {journal} {Phys. Rev. Lett.}\ }\textbf {\bibinfo {volume} {114}},\
  \bibinfo {pages} {103001} (\bibinfo {year} {2015})}\BibitemShut {NoStop}%
\bibitem [{\citenamefont {Prost}\ \emph {et~al.}(2017)\citenamefont {Prost},
  \citenamefont {Zhang}, \citenamefont {Hertz}, \citenamefont {Billard},
  \citenamefont {Lavorel}, \citenamefont {Bejot}, \citenamefont {Zyss},
  \citenamefont {{Sh. Averbukh}},\ and\ \citenamefont
  {Faucher}}]{Prost2017Third}%
  \BibitemOpen
  \bibfield  {author} {\bibinfo {author} {\bibfnamefont {E.}~\bibnamefont
  {Prost}}, \bibinfo {author} {\bibfnamefont {H.}~\bibnamefont {Zhang}},
  \bibinfo {author} {\bibfnamefont {E.}~\bibnamefont {Hertz}}, \bibinfo
  {author} {\bibfnamefont {F.}~\bibnamefont {Billard}}, \bibinfo {author}
  {\bibfnamefont {B.}~\bibnamefont {Lavorel}}, \bibinfo {author} {\bibfnamefont
  {P.}~\bibnamefont {Bejot}}, \bibinfo {author} {\bibfnamefont
  {J.}~\bibnamefont {Zyss}}, \bibinfo {author} {\bibfnamefont {I.}~\bibnamefont
  {{Sh. Averbukh}}},\ and\ \bibinfo {author} {\bibfnamefont {O.}~\bibnamefont
  {Faucher}},\ }\bibfield  {title} {\bibinfo {title} {Third-order-harmonic
  generation in coherently spinning molecules},\ }\href
  {https://doi.org/10.1103/PhysRevA.96.043418} {\bibfield  {journal} {\bibinfo
  {journal} {Phys. Rev. A}\ }\textbf {\bibinfo {volume} {96}},\ \bibinfo
  {pages} {043418} (\bibinfo {year} {2017})}\BibitemShut {NoStop}%
\bibitem [{\citenamefont {Mizuse}\ \emph {et~al.}(2020)\citenamefont {Mizuse},
  \citenamefont {Sakamoto}, \citenamefont {Fujimoto},\ and\ \citenamefont
  {Ohshima}}]{Mizuse2020Direct}%
  \BibitemOpen
  \bibfield  {author} {\bibinfo {author} {\bibfnamefont {K.}~\bibnamefont
  {Mizuse}}, \bibinfo {author} {\bibfnamefont {N.}~\bibnamefont {Sakamoto}},
  \bibinfo {author} {\bibfnamefont {R.}~\bibnamefont {Fujimoto}},\ and\
  \bibinfo {author} {\bibfnamefont {Y.}~\bibnamefont {Ohshima}},\ }\bibfield
  {title} {\bibinfo {title} {Direct imaging of direction-controlled molecular
  rotational wave packets created by a polarization-skewed double-pulse},\
  }\href {https://doi.org/10.1039/D0CP01084G} {\bibfield  {journal} {\bibinfo
  {journal} {Phys. Chem. Chem. Phys.}\ }\textbf {\bibinfo {volume} {22}},\
  \bibinfo {pages} {10853} (\bibinfo {year} {2020})}\BibitemShut {NoStop}%
\bibitem [{\citenamefont {Tutunnikov}\ \emph {et~al.}(2021)\citenamefont
  {Tutunnikov}, \citenamefont {Xu}, \citenamefont {Field}, \citenamefont
  {Nelson}, \citenamefont {Prior},\ and\ \citenamefont {{Sh.
  Averbukh}}}]{Tutunnikov2021Enantioselective}%
  \BibitemOpen
  \bibfield  {author} {\bibinfo {author} {\bibfnamefont {I.}~\bibnamefont
  {Tutunnikov}}, \bibinfo {author} {\bibfnamefont {L.}~\bibnamefont {Xu}},
  \bibinfo {author} {\bibfnamefont {R.~W.}\ \bibnamefont {Field}}, \bibinfo
  {author} {\bibfnamefont {K.~A.}\ \bibnamefont {Nelson}}, \bibinfo {author}
  {\bibfnamefont {Y.}~\bibnamefont {Prior}},\ and\ \bibinfo {author}
  {\bibfnamefont {I.}~\bibnamefont {{Sh. Averbukh}}},\ }\bibfield  {title}
  {\bibinfo {title} {Enantioselective orientation of chiral molecules induced
  by terahertz pulses with twisted polarization},\ }\href
  {https://doi.org/10.1103/PhysRevResearch.3.013249} {\bibfield  {journal}
  {\bibinfo  {journal} {Phys. Rev. Research}\ }\textbf {\bibinfo {volume}
  {3}},\ \bibinfo {pages} {013249} (\bibinfo {year} {2021})}\BibitemShut
  {NoStop}%
\bibitem [{\citenamefont {Takemoto}\ and\ \citenamefont
  {Yamanouchi}(2008)}]{Takemoto2008Fixing}%
  \BibitemOpen
  \bibfield  {author} {\bibinfo {author} {\bibfnamefont {N.}~\bibnamefont
  {Takemoto}}\ and\ \bibinfo {author} {\bibfnamefont {K.}~\bibnamefont
  {Yamanouchi}},\ }\bibfield  {title} {\bibinfo {title} {Fixing chiral
  molecules in space by intense two-color phase-locked laser fields},\ }\href
  {https://doi.org/https://doi.org/10.1016/j.cplett.2007.11.037} {\bibfield
  {journal} {\bibinfo  {journal} {Chem. Phys. Lett.}\ }\textbf {\bibinfo
  {volume} {451}},\ \bibinfo {pages} {1 } (\bibinfo {year} {2008})}\BibitemShut
  {NoStop}%
\bibitem [{\citenamefont {Xu}\ \emph {et~al.}(2021)\citenamefont {Xu},
  \citenamefont {Tutunnikov}, \citenamefont {Prior},\ and\ \citenamefont {{Sh.
  Averbukh}}}]{Xu2021Three}%
  \BibitemOpen
  \bibfield  {author} {\bibinfo {author} {\bibfnamefont {L.}~\bibnamefont
  {Xu}}, \bibinfo {author} {\bibfnamefont {I.}~\bibnamefont {Tutunnikov}},
  \bibinfo {author} {\bibfnamefont {Y.}~\bibnamefont {Prior}},\ and\ \bibinfo
  {author} {\bibfnamefont {I.}~\bibnamefont {{Sh. Averbukh}}},\ }\bibfield
  {title} {\bibinfo {title} {Three dimensional orientation of small polyatomic
  molecules excited by two-color femtosecond pulses},\ }\href
  {https://doi.org/10.1088/1361-6455/ac20e3} {\bibfield  {journal} {\bibinfo
  {journal} {J Phys. B}\ }\textbf {\bibinfo {volume} {54}},\ \bibinfo {pages}
  {164003} (\bibinfo {year} {2021})}\BibitemShut {NoStop}%
\bibitem [{\citenamefont {Korech}\ \emph {et~al.}(2013)\citenamefont {Korech},
  \citenamefont {Steinitz}, \citenamefont {Gordon}, \citenamefont {{Sh.
  Averbukh}},\ and\ \citenamefont {Prior}}]{Korech2013Observing}%
  \BibitemOpen
  \bibfield  {author} {\bibinfo {author} {\bibfnamefont {O.}~\bibnamefont
  {Korech}}, \bibinfo {author} {\bibfnamefont {U.}~\bibnamefont {Steinitz}},
  \bibinfo {author} {\bibfnamefont {R.~J.}\ \bibnamefont {Gordon}}, \bibinfo
  {author} {\bibfnamefont {I.}~\bibnamefont {{Sh. Averbukh}}},\ and\ \bibinfo
  {author} {\bibfnamefont {Y.}~\bibnamefont {Prior}},\ }\bibfield  {title}
  {\bibinfo {title} {Observing molecular spinning via the rotational doppler
  effect},\ }\href {https://doi.org/10.1038/nphoton.2013.189} {\bibfield
  {journal} {\bibinfo  {journal} {Nat. Photon.}\ }\textbf {\bibinfo {volume}
  {7}},\ \bibinfo {pages} {711} (\bibinfo {year} {2013})}\BibitemShut {NoStop}%
\bibitem [{\citenamefont {Mizuse}\ \emph {et~al.}(2015)\citenamefont {Mizuse},
  \citenamefont {Kitano}, \citenamefont {Hasegawa},\ and\ \citenamefont
  {Ohshima}}]{Mizuse2015Quantum}%
  \BibitemOpen
  \bibfield  {author} {\bibinfo {author} {\bibfnamefont {K.}~\bibnamefont
  {Mizuse}}, \bibinfo {author} {\bibfnamefont {K.}~\bibnamefont {Kitano}},
  \bibinfo {author} {\bibfnamefont {H.}~\bibnamefont {Hasegawa}},\ and\
  \bibinfo {author} {\bibfnamefont {Y.}~\bibnamefont {Ohshima}},\ }\bibfield
  {title} {\bibinfo {title} {Quantum unidirectional rotation directly imaged
  with molecules},\ }\href {https://doi.org/10.1126/sciadv.1400185} {\bibfield
  {journal} {\bibinfo  {journal} {Sci. Adv.}\ }\textbf {\bibinfo {volume}
  {1}},\ \bibinfo {pages} {e1400185} (\bibinfo {year} {2015})}\BibitemShut
  {NoStop}%
\bibitem [{\citenamefont {Lin}\ \emph {et~al.}(2015)\citenamefont {Lin},
  \citenamefont {Song}, \citenamefont {Gong}, \citenamefont {Ji}, \citenamefont
  {Pan}, \citenamefont {Ding}, \citenamefont {Zeng},\ and\ \citenamefont
  {Wu}}]{Lin2015Visualizing}%
  \BibitemOpen
  \bibfield  {author} {\bibinfo {author} {\bibfnamefont {K.}~\bibnamefont
  {Lin}}, \bibinfo {author} {\bibfnamefont {Q.}~\bibnamefont {Song}}, \bibinfo
  {author} {\bibfnamefont {X.}~\bibnamefont {Gong}}, \bibinfo {author}
  {\bibfnamefont {Q.}~\bibnamefont {Ji}}, \bibinfo {author} {\bibfnamefont
  {H.}~\bibnamefont {Pan}}, \bibinfo {author} {\bibfnamefont {J.}~\bibnamefont
  {Ding}}, \bibinfo {author} {\bibfnamefont {H.}~\bibnamefont {Zeng}},\ and\
  \bibinfo {author} {\bibfnamefont {J.}~\bibnamefont {Wu}},\ }\bibfield
  {title} {\bibinfo {title} {Visualizing molecular unidirectional rotation},\
  }\href {https://doi.org/10.1103/PhysRevA.92.013410} {\bibfield  {journal}
  {\bibinfo  {journal} {Phys. Rev. A}\ }\textbf {\bibinfo {volume} {92}},\
  \bibinfo {pages} {013410} (\bibinfo {year} {2015})}\BibitemShut {NoStop}%
\bibitem [{\citenamefont {Seideman}(2001)}]{Seideman2001On}%
  \BibitemOpen
  \bibfield  {author} {\bibinfo {author} {\bibfnamefont {T.}~\bibnamefont
  {Seideman}},\ }\bibfield  {title} {\bibinfo {title} {On the dynamics of
  rotationally broad, spatially aligned wave packets},\ }\href
  {https://doi.org/10.1063/1.1400131} {\bibfield  {journal} {\bibinfo
  {journal} {J. Chem. Phys.}\ }\textbf {\bibinfo {volume} {115}},\ \bibinfo
  {pages} {5965} (\bibinfo {year} {2001})}\BibitemShut {NoStop}%
\bibitem [{\citenamefont {Underwood}\ \emph {et~al.}(2003)\citenamefont
  {Underwood}, \citenamefont {Spanner}, \citenamefont {{Yu. Ivanov}},
  \citenamefont {Mottershead}, \citenamefont {Sussman},\ and\ \citenamefont
  {Stolow}}]{Underwood2003Switched}%
  \BibitemOpen
  \bibfield  {author} {\bibinfo {author} {\bibfnamefont {J.~G.}\ \bibnamefont
  {Underwood}}, \bibinfo {author} {\bibfnamefont {M.}~\bibnamefont {Spanner}},
  \bibinfo {author} {\bibfnamefont {M.}~\bibnamefont {{Yu. Ivanov}}}, \bibinfo
  {author} {\bibfnamefont {J.}~\bibnamefont {Mottershead}}, \bibinfo {author}
  {\bibfnamefont {B.~J.}\ \bibnamefont {Sussman}},\ and\ \bibinfo {author}
  {\bibfnamefont {A.}~\bibnamefont {Stolow}},\ }\bibfield  {title} {\bibinfo
  {title} {Switched wave packets: A route to nonperturbative quantum control},\
  }\href {https://doi.org/10.1103/PhysRevLett.90.223001} {\bibfield  {journal}
  {\bibinfo  {journal} {Phys. Rev. Lett.}\ }\textbf {\bibinfo {volume} {90}},\
  \bibinfo {pages} {223001} (\bibinfo {year} {2003})}\BibitemShut {NoStop}%
\bibitem [{\citenamefont {Underwood}\ \emph {et~al.}(2005)\citenamefont
  {Underwood}, \citenamefont {Sussman},\ and\ \citenamefont
  {Stolow}}]{Underwood2005Field}%
  \BibitemOpen
  \bibfield  {author} {\bibinfo {author} {\bibfnamefont {J.~G.}\ \bibnamefont
  {Underwood}}, \bibinfo {author} {\bibfnamefont {B.~J.}\ \bibnamefont
  {Sussman}},\ and\ \bibinfo {author} {\bibfnamefont {A.}~\bibnamefont
  {Stolow}},\ }\bibfield  {title} {\bibinfo {title} {Field-free three
  dimensional molecular axis alignment},\ }\href
  {https://doi.org/10.1103/PhysRevLett.94.143002} {\bibfield  {journal}
  {\bibinfo  {journal} {Phys. Rev. Lett.}\ }\textbf {\bibinfo {volume} {94}},\
  \bibinfo {pages} {143002} (\bibinfo {year} {2005})}\BibitemShut {NoStop}%
\bibitem [{\citenamefont {Goban}\ \emph {et~al.}(2008)\citenamefont {Goban},
  \citenamefont {Minemoto},\ and\ \citenamefont {Sakai}}]{Goban2008Laser}%
  \BibitemOpen
  \bibfield  {author} {\bibinfo {author} {\bibfnamefont {A.}~\bibnamefont
  {Goban}}, \bibinfo {author} {\bibfnamefont {S.}~\bibnamefont {Minemoto}},\
  and\ \bibinfo {author} {\bibfnamefont {H.}~\bibnamefont {Sakai}},\ }\bibfield
   {title} {\bibinfo {title} {Laser-field-free molecular orientation},\ }\href
  {https://doi.org/10.1103/PhysRevLett.101.013001} {\bibfield  {journal}
  {\bibinfo  {journal} {Phys. Rev. Lett.}\ }\textbf {\bibinfo {volume} {101}},\
  \bibinfo {pages} {013001} (\bibinfo {year} {2008})}\BibitemShut {NoStop}%
\bibitem [{\citenamefont {Chatterley}\ \emph {et~al.}(2019)\citenamefont
  {Chatterley}, \citenamefont {Schouder}, \citenamefont {Christiansen},
  \citenamefont {Shepperson}, \citenamefont {Rasmussen},\ and\ \citenamefont
  {Stapelfeldt}}]{Chatterley2019Long}%
  \BibitemOpen
  \bibfield  {author} {\bibinfo {author} {\bibfnamefont {A.~S.}\ \bibnamefont
  {Chatterley}}, \bibinfo {author} {\bibfnamefont {C.}~\bibnamefont
  {Schouder}}, \bibinfo {author} {\bibfnamefont {L.}~\bibnamefont
  {Christiansen}}, \bibinfo {author} {\bibfnamefont {B.}~\bibnamefont
  {Shepperson}}, \bibinfo {author} {\bibfnamefont {M.~H.}\ \bibnamefont
  {Rasmussen}},\ and\ \bibinfo {author} {\bibfnamefont {H.}~\bibnamefont
  {Stapelfeldt}},\ }\bibfield  {title} {\bibinfo {title} {Long-lasting
  field-free alignment of large molecules inside helium nanodroplets},\ }\href
  {https://doi.org/10.1038/s41467-018-07995-0} {\bibfield  {journal} {\bibinfo
  {journal} {Nat. Commun.}\ }\textbf {\bibinfo {volume} {10}},\ \bibinfo
  {pages} {133} (\bibinfo {year} {2019})}\BibitemShut {NoStop}%
\bibitem [{\citenamefont {Krems}(2018)}]{Krems2018Molecules}%
  \BibitemOpen
  \bibfield  {author} {\bibinfo {author} {\bibfnamefont {R.~V.}\ \bibnamefont
  {Krems}},\ }\href@noop {} {\emph {\bibinfo {title} {Molecules in
  Electromagnetic Fields: From Ultracold Physics to Controlled Chemistry}}}\
  (\bibinfo  {publisher} {Wiley, New York},\ \bibinfo {year}
  {2018})\BibitemShut {NoStop}%
\bibitem [{\citenamefont {Koch}\ \emph {et~al.}(2019)\citenamefont {Koch},
  \citenamefont {Lemeshko},\ and\ \citenamefont {Sugny}}]{Koch2019Quantum}%
  \BibitemOpen
  \bibfield  {author} {\bibinfo {author} {\bibfnamefont {C.~P.}\ \bibnamefont
  {Koch}}, \bibinfo {author} {\bibfnamefont {M.}~\bibnamefont {Lemeshko}},\
  and\ \bibinfo {author} {\bibfnamefont {D.}~\bibnamefont {Sugny}},\ }\bibfield
   {title} {\bibinfo {title} {Quantum control of molecular rotation},\ }\href
  {https://doi.org/10.1103/RevModPhys.91.035005} {\bibfield  {journal}
  {\bibinfo  {journal} {Rev. Mod. Phys.}\ }\textbf {\bibinfo {volume} {91}},\
  \bibinfo {pages} {035005} (\bibinfo {year} {2019})}\BibitemShut {NoStop}%
\bibitem [{\citenamefont {Zare}(1988)}]{zare1988Angular}%
  \BibitemOpen
  \bibfield  {author} {\bibinfo {author} {\bibfnamefont {R.~N.}\ \bibnamefont
  {Zare}},\ }\href@noop {} {\emph {\bibinfo {title} {Angular momentum:
  understanding spatial aspects in chemistry and physics}}}\ (\bibinfo
  {publisher} {Wiley, New York},\ \bibinfo {year} {1988})\BibitemShut {NoStop}%
\bibitem [{\citenamefont {Sidje}(1998)}]{sidje1998Expokit}%
  \BibitemOpen
  \bibfield  {author} {\bibinfo {author} {\bibfnamefont {R.~B.}\ \bibnamefont
  {Sidje}},\ }\bibfield  {title} {\bibinfo {title} {Expokit: {A} software
  package for computing matrix exponentials},\ }\href
  {https://doi.org/10.1145/285861.285868} {\bibfield  {journal} {\bibinfo
  {journal} {ACM Trans. Math. Softw.}\ }\textbf {\bibinfo {volume} {24}},\
  \bibinfo {pages} {130} (\bibinfo {year} {1998})}\BibitemShut {NoStop}%
\bibitem [{\citenamefont {Frisch}\ \emph {et~al.}(2016)\citenamefont {Frisch},
  \citenamefont {Trucks}, \citenamefont {Schlegel}, \citenamefont {Scuseria},
  \citenamefont {Robb}, \citenamefont {Cheeseman}, \citenamefont {Scalmani},
  \citenamefont {Barone}, \citenamefont {Petersson}, \citenamefont {Nakatsuji},
  \citenamefont {Li}, \citenamefont {Caricato}, \citenamefont {Marenich},
  \citenamefont {Bloino}, \citenamefont {Janesko}, \citenamefont {Gomperts},
  \citenamefont {Mennucci}, \citenamefont {Hratchian}, \citenamefont {Ortiz},
  \citenamefont {Izmaylov} \emph {et~al.}}]{Frisch2016Gaussian}%
  \BibitemOpen
  \bibfield  {author} {\bibinfo {author} {\bibfnamefont {M.~J.}\ \bibnamefont
  {Frisch}}, \bibinfo {author} {\bibfnamefont {G.~W.}\ \bibnamefont {Trucks}},
  \bibinfo {author} {\bibfnamefont {H.~B.}\ \bibnamefont {Schlegel}}, \bibinfo
  {author} {\bibfnamefont {G.~E.}\ \bibnamefont {Scuseria}}, \bibinfo {author}
  {\bibfnamefont {M.~A.}\ \bibnamefont {Robb}}, \bibinfo {author}
  {\bibfnamefont {J.~R.}\ \bibnamefont {Cheeseman}}, \bibinfo {author}
  {\bibfnamefont {G.}~\bibnamefont {Scalmani}}, \bibinfo {author}
  {\bibfnamefont {V.}~\bibnamefont {Barone}}, \bibinfo {author} {\bibfnamefont
  {G.~A.}\ \bibnamefont {Petersson}}, \bibinfo {author} {\bibfnamefont
  {H.}~\bibnamefont {Nakatsuji}}, \bibinfo {author} {\bibfnamefont
  {X.}~\bibnamefont {Li}}, \bibinfo {author} {\bibfnamefont {M.}~\bibnamefont
  {Caricato}}, \bibinfo {author} {\bibfnamefont {A.~V.}\ \bibnamefont
  {Marenich}}, \bibinfo {author} {\bibfnamefont {J.}~\bibnamefont {Bloino}},
  \bibinfo {author} {\bibfnamefont {B.~G.}\ \bibnamefont {Janesko}}, \bibinfo
  {author} {\bibfnamefont {R.}~\bibnamefont {Gomperts}}, \bibinfo {author}
  {\bibfnamefont {B.}~\bibnamefont {Mennucci}}, \bibinfo {author}
  {\bibfnamefont {H.~P.}\ \bibnamefont {Hratchian}}, \bibinfo {author}
  {\bibfnamefont {J.~V.}\ \bibnamefont {Ortiz}}, \bibinfo {author}
  {\bibfnamefont {A.~F.}\ \bibnamefont {Izmaylov}}, \emph {et~al.},\
  }\href@noop {} {\bibinfo {title} {Gaussian~16 {R}evision {A}. 03, {Gaussian
  Inc. Wallingford CT}}} (\bibinfo {year} {2016})\BibitemShut {NoStop}%
\bibitem [{\citenamefont {Goldstein}\ \emph {et~al.}(2002)\citenamefont
  {Goldstein}, \citenamefont {Poole},\ and\ \citenamefont
  {Safko}}]{Goldstein2002Classical}%
  \BibitemOpen
  \bibfield  {author} {\bibinfo {author} {\bibfnamefont {H.}~\bibnamefont
  {Goldstein}}, \bibinfo {author} {\bibfnamefont {C.}~\bibnamefont {Poole}},\
  and\ \bibinfo {author} {\bibfnamefont {J.}~\bibnamefont {Safko}},\
  }\href@noop {} {\emph {\bibinfo {title} {{Classical Mechanics}}}}\ (\bibinfo
  {publisher} {Addison Wesley, San Francisco, CA},\ \bibinfo {year}
  {2002})\BibitemShut {NoStop}%
\bibitem [{\citenamefont {{Sh. Averbukh}}\ and\ \citenamefont
  {Arvieu}(2001)}]{Averbukh2001}%
  \BibitemOpen
  \bibfield  {author} {\bibinfo {author} {\bibfnamefont {I.}~\bibnamefont {{Sh.
  Averbukh}}}\ and\ \bibinfo {author} {\bibfnamefont {R.}~\bibnamefont
  {Arvieu}},\ }\bibfield  {title} {\bibinfo {title} {Angular focusing,
  squeezing, and rainbow formation in a strongly driven quantum rotor},\ }\href
  {https://doi.org/10.1103/PhysRevLett.87.163601} {\bibfield  {journal}
  {\bibinfo  {journal} {Phys. Rev. Lett.}\ }\textbf {\bibinfo {volume} {87}},\
  \bibinfo {pages} {163601} (\bibinfo {year} {2001})}\BibitemShut {NoStop}%
\bibitem [{\citenamefont {Leibscher}\ \emph {et~al.}(2003)\citenamefont
  {Leibscher}, \citenamefont {{Sh. Averbukh}},\ and\ \citenamefont
  {Rabitz}}]{Averbukh2003}%
  \BibitemOpen
  \bibfield  {author} {\bibinfo {author} {\bibfnamefont {M.}~\bibnamefont
  {Leibscher}}, \bibinfo {author} {\bibfnamefont {I.}~\bibnamefont {{Sh.
  Averbukh}}},\ and\ \bibinfo {author} {\bibfnamefont {H.}~\bibnamefont
  {Rabitz}},\ }\bibfield  {title} {\bibinfo {title} {Molecular alignment by
  trains of short laser pulses},\ }\href
  {https://doi.org/10.1103/PhysRevLett.90.213001} {\bibfield  {journal}
  {\bibinfo  {journal} {Phys. Rev. Lett.}\ }\textbf {\bibinfo {volume} {90}},\
  \bibinfo {pages} {213001} (\bibinfo {year} {2003})}\BibitemShut {NoStop}%
\bibitem [{\citenamefont {Leibscher}\ \emph {et~al.}(2004)\citenamefont
  {Leibscher}, \citenamefont {{Sh. Averbukh}},\ and\ \citenamefont
  {Rabitz}}]{Averbukh2004}%
  \BibitemOpen
  \bibfield  {author} {\bibinfo {author} {\bibfnamefont {M.}~\bibnamefont
  {Leibscher}}, \bibinfo {author} {\bibfnamefont {I.}~\bibnamefont {{Sh.
  Averbukh}}},\ and\ \bibinfo {author} {\bibfnamefont {H.}~\bibnamefont
  {Rabitz}},\ }\bibfield  {title} {\bibinfo {title} {Enhanced molecular
  alignment by short laser pulses},\ }\href
  {https://doi.org/10.1103/PhysRevA.69.013402} {\bibfield  {journal} {\bibinfo
  {journal} {Phys. Rev. A}\ }\textbf {\bibinfo {volume} {69}},\ \bibinfo
  {pages} {013402} (\bibinfo {year} {2004})}\BibitemShut {NoStop}%
\bibitem [{\citenamefont {Lee}\ \emph {et~al.}(2004)\citenamefont {Lee},
  \citenamefont {Litvinyuk}, \citenamefont {Dooley}, \citenamefont {Spanner},
  \citenamefont {Villeneuve},\ and\ \citenamefont {Corkum}}]{Lee2004}%
  \BibitemOpen
  \bibfield  {author} {\bibinfo {author} {\bibfnamefont {K.~F.}\ \bibnamefont
  {Lee}}, \bibinfo {author} {\bibfnamefont {I.~V.}\ \bibnamefont {Litvinyuk}},
  \bibinfo {author} {\bibfnamefont {P.~W.}\ \bibnamefont {Dooley}}, \bibinfo
  {author} {\bibfnamefont {M.}~\bibnamefont {Spanner}}, \bibinfo {author}
  {\bibfnamefont {D.~M.}\ \bibnamefont {Villeneuve}},\ and\ \bibinfo {author}
  {\bibfnamefont {P.~B.}\ \bibnamefont {Corkum}},\ }\bibfield  {title}
  {\bibinfo {title} {Two-pulse alignment of molecules},\ }\href
  {http://stacks.iop.org/0953-4075/37/i=3/a=L02} {\bibfield  {journal}
  {\bibinfo  {journal} {J. Phys. B}\ }\textbf {\bibinfo {volume} {37}},\
  \bibinfo {pages} {L43} (\bibinfo {year} {2004})}\BibitemShut {NoStop}%
\bibitem [{\citenamefont {Bisgaard}\ \emph {et~al.}(2004)\citenamefont
  {Bisgaard}, \citenamefont {Poulsen}, \citenamefont {P\'eronne}, \citenamefont
  {Viftrup},\ and\ \citenamefont {Stapelfeldt}}]{Bisgaard2004}%
  \BibitemOpen
  \bibfield  {author} {\bibinfo {author} {\bibfnamefont {C.~Z.}\ \bibnamefont
  {Bisgaard}}, \bibinfo {author} {\bibfnamefont {M.~D.}\ \bibnamefont
  {Poulsen}}, \bibinfo {author} {\bibfnamefont {E.}~\bibnamefont {P\'eronne}},
  \bibinfo {author} {\bibfnamefont {S.~S.}\ \bibnamefont {Viftrup}},\ and\
  \bibinfo {author} {\bibfnamefont {H.}~\bibnamefont {Stapelfeldt}},\
  }\bibfield  {title} {\bibinfo {title} {Observation of enhanced field-free
  molecular alignment by two laser pulses},\ }\href
  {https://doi.org/10.1103/PhysRevLett.92.173004} {\bibfield  {journal}
  {\bibinfo  {journal} {Phys. Rev. Lett.}\ }\textbf {\bibinfo {volume} {92}},\
  \bibinfo {pages} {173004} (\bibinfo {year} {2004})}\BibitemShut {NoStop}%
\bibitem [{\citenamefont {Pinkham}\ \emph {et~al.}(2007)\citenamefont
  {Pinkham}, \citenamefont {Mooney},\ and\ \citenamefont
  {Jones}}]{Pinkham2007}%
  \BibitemOpen
  \bibfield  {author} {\bibinfo {author} {\bibfnamefont {D.}~\bibnamefont
  {Pinkham}}, \bibinfo {author} {\bibfnamefont {K.~E.}\ \bibnamefont
  {Mooney}},\ and\ \bibinfo {author} {\bibfnamefont {R.~R.}\ \bibnamefont
  {Jones}},\ }\bibfield  {title} {\bibinfo {title} {Optimizing dynamic
  alignment in room temperature {CO}},\ }\href
  {https://doi.org/10.1103/PhysRevA.75.013422} {\bibfield  {journal} {\bibinfo
  {journal} {Phys. Rev. A}\ }\textbf {\bibinfo {volume} {75}},\ \bibinfo
  {pages} {013422} (\bibinfo {year} {2007})}\BibitemShut {NoStop}%
\bibitem [{\citenamefont {Yachmenev}\ \emph {et~al.}(2019)\citenamefont
  {Yachmenev}, \citenamefont {Onvlee}, \citenamefont {Zak}, \citenamefont
  {Owens},\ and\ \citenamefont {K\"upper}}]{Yachmenev2019Field}%
  \BibitemOpen
  \bibfield  {author} {\bibinfo {author} {\bibfnamefont {A.}~\bibnamefont
  {Yachmenev}}, \bibinfo {author} {\bibfnamefont {J.}~\bibnamefont {Onvlee}},
  \bibinfo {author} {\bibfnamefont {E.}~\bibnamefont {Zak}}, \bibinfo {author}
  {\bibfnamefont {A.}~\bibnamefont {Owens}},\ and\ \bibinfo {author}
  {\bibfnamefont {J.}~\bibnamefont {K\"upper}},\ }\bibfield  {title} {\bibinfo
  {title} {{Field-Induced Diastereomers for Chiral Separation}},\ }\href
  {https://doi.org/10.1103/PhysRevLett.123.243202} {\bibfield  {journal}
  {\bibinfo  {journal} {Phys. Rev. Lett.}\ }\textbf {\bibinfo {volume} {123}},\
  \bibinfo {pages} {243202} (\bibinfo {year} {2019})}\BibitemShut {NoStop}%
\end{thebibliography}%

\end{document}